\documentclass[english,aps, twocolumn, pre,superscriptaddress, notitlepage]{revtex4-1}
\usepackage[LGR,T1]{fontenc}
\usepackage[latin9]{inputenc}
\usepackage{verbatim}
\usepackage{amsmath}
\usepackage{amssymb}
\usepackage{graphicx}

\makeatletter

\DeclareRobustCommand{\greektext}{%
  \fontencoding{LGR}\selectfont\def\encodingdefault{LGR}}
\DeclareRobustCommand{\textgreek}[1]{\leavevmode{\greektext #1}}
\ProvideTextCommand{\~}{LGR}[1]{\char126#1}

\providecommand{\tabularnewline}{\\}

\usepackage[colorlinks=true,linktocpage=true,linkcolor=MidnightBlue,urlcolor=MidnightBlue,citecolor=MidnightBlue,anchorcolor=MidnightBlue]{hyperref}
\usepackage[dvipsnames]{xcolor}
\usepackage{mhchem}
\renewcommand{\textendash}{--}

\makeatother

\usepackage{babel}
\begin{document}

\title{Transients generate memory and break hyperbolicity in stochastic
enzymatic networks}

\author{Ashutosh Kumar}

\affiliation{Department of Chemistry, Indian Institute of Technology, Madras,
Chennai 600036, India}

\author{R. Adhikari}

\affiliation{DAMTP, Centre for Mathematical Sciences, University of Cambridge,
Wilberforce Road, Cambridge CB3 0WA, UK}

\author{Arti Dua}

\affiliation{Department of Chemistry, Indian Institute of Technology, Madras,
Chennai 600036, India}
\begin{abstract}
The hyperbolic dependence of catalytic rate on substrate concentration
is a classical result in enzyme kinetics, quantified by the celebrated
Michaelis-Menten equation. The ubiquity of this relation in diverse
chemical and biological contexts has recently been rationalized by
a graph-theoretic analysis of deterministic reaction networks. Experiments,
however, have revealed that ``molecular noise'' - intrinsic stochasticity
at the molecular scale - leads to significant deviations from classical
results and to unexpected effects like ``molecular memory'', i.e.,
the breakdown of statistical independence between turnover events.
Here we show, through a new method of analysis, that memory and non-hyperbolicity
have a common source in an initial, and observably long, transient
peculiar to stochastic reaction networks of multiple enzymes. Networks
of single enzymes do not admit such transients. The transient yields,
asymptotically, to a steady-state in which memory vanishes and hyperbolicity
is recovered. We propose new statistical measures, defined in terms
of turnover times, to distinguish between the transient and steady
states and apply these to experimental data from a landmark experiment
that first observed molecular memory in a single enzyme with multiple
binding sites. Our study shows that catalysis at the molecular level
with more than one enzyme always contains a non-classical regime and
provides insight on how the classical limit is attained.
\end{abstract}
\maketitle

\section{Introduction}

The Michaelis-Menten equation, describing the hyperbolic dependence
of the rate of catalysis on the substrate concentration, is a classical
result in enzyme kinetics \cite{key-1}. It was derived by Michaelis
and Menten in 1913 for a network of three elementary reactions, $E+S\rightleftharpoons ES\rightarrow E+P,$
describing the reversible binding of enzyme $E$ with substrate $S$
to form complex $ES$ and its irreversible dissociation into product
$P$ and regenerated enzyme $E$ \cite{key-2,key-3,key-4}. The hyperbolic
dependence of catalytic rate on substrate concentration is found to
hold in enzymatic networks of far greater complexity. It implies a
linear relation between the inverse catalytic rate and the inverse
substrate concentration and, in this form, is widely used to estimate
rate parameters and infer mechanisms from kinetic data \cite{key-8}.
The surprising ubiquity of this equation in chemical and biological
processes has recently been rationalized by a graph-theoretical analysis
of complex, deterministic, reaction networks \cite{key-9}. 

At the molecular level, however, enzymatic reactions do not proceed
deterministically \cite{key-10,key-11,key-12,key-13,key-14}. Fluctuations,
of both quantum mechanical and thermal origin, termed ``molecular
noise'', influence each step of a chemical reaction, such that neither
the lifetime of a chemical state nor the state to which it transits
can be known with certainty \cite{key-15,key-16,key-17}. Further,
the discrete change in the reactant numbers is comparable to the number
of reacting molecules, and a description in terms of continuously
varying concentrations is inadmissible \cite{key-15,key-16,key-17,key-58,key-18}.
In the limit of large numbers of reactants, when both fluctuations
and the change in reactants compared to their total number are small,
a deterministic description in terms of continuously varying concentrations
is recovered \cite{key-18}. The Michaelis-Menten equation is obtained
when, in addition, there is a separation of time scales between the
(rapid) equilibration between enzyme and complex and (slow) product
formation \cite{key-2}. This rapid equilibrium approximation is a
special case of the steady-state approximation (SSA), in which the
rates of complex formation and dissociation are assumed to be equal,
as noted by Briggs and Haldane\textbf{ }\cite{key-56}.\textbf{ }

The first theoretical study of catalytic fluctuations was undertaken
by Bartholomay half a century after the discovery of the Michaelis-Menten
(MM) equation \cite{key-18}. His principal contribution was to show
that discrete-state continuous-time Markov processes provide a mathematical
framework that incorporates the discrete change in molecular numbers,
the effect of molecular noise in each reaction step, and reactions
mechanisms of arbitrary complexity. The classical rate equations for
concentrations were thus replaced by chemical ``master equations''
for the probabilities of the (non-negative) number of reactants. Bartholomay
obtained the mean and variance of these for the Michaelis-Menten mechanism
$E+S\rightleftharpoons ES\rightarrow E+P$. The apparent irreproducibility
of experiments that measured the rate of change of concentrations
was recognized to be a fluctuation effect and a method was suggested
to estimate the rate constraints from the variances of the concentrations. 

The long hiatus of interest that followed this pioneering work was
brought to a close by a landmark experiment that directly observed
catalytic fluctuations at the single-molecule level \cite{key-12,key-13}.
As concentrations are not defined for a single molecule, the experiment
measured, instead, the times at which the enzyme yielded products,
one product at a time. This time series data was analyzed in terms
of the interval between consecutive turnovers, defined to be the ``waiting
time''. For repeated experiments under identical conditions, the
waiting times showed a distribution and this was attributed to the
effect of molecular noise. The analysis of waiting time distributions
revealed several remarkable facts. First, the distribution changed
character with increase in substrate concentration, from a single
exponential to one that was not. Second, the inverse of the mean waiting
time obeyed the MME at low substrate concentrations. Third, the randomness
parameter, the ratio of the variance to the squared mean, was a monotonically
increasing function of the substrate concentration, bounded below
by one. Fourth, the waiting times between consecutive turnovers were
found to be statistically dependent, with substantial positive correlations,
an effect termed ``molecular memory''. Subsequent experiments in
single-nanoparticle catalysis confirmed these empirical facts and
established their generality \cite{key-57,key-55,key-20}. While it
was understood that these seemingly disparate observations have their
origin in molecular noise, the precise manner in which they emerge
from underlying molecular fluctuations and how they are influenced
by different reaction mechanisms was not elucidated. 

The central theoretical question that needs to be answered in rationalizing
such single-molecule temporal data is this: can we derive the statistics
of \emph{temporal} fluctuations from the chemical master equation,
incorporating discreteness, molecular noise, and reaction mechanisms,
in the manner that the statistics of \emph{number }fluctuations was
derived by Bartholomay? Here we present a formalism that permits us
to answer this question affirmatively. Using this formalism we are
able to make a direct connection between reaction mechanisms and the
statistics of waiting times and, thus, explain their puzzling features
from a unified point of view. 

In Section \ref{sec:Stochastic-enzymatic-networks} we consider a
generic stochastic enzymatic reaction network, incorporating conformational
fluctuations and parallel pathways to product formation, and present
the corresponding chemical master equation (CME). We marginalize the
reactant probabilities to obtain the probability of there being $n$
turnovers at any given time and present several experimentally relevant
summary statistics. We introduce the probability distributions of
the turnover and waiting times and present their relevant summary
statistics. We then derive an expression that connects the reactant
probabilities of the CME to the distribution of waiting times. This
provides the sought after link between the description in terms of
waiting times (``point process'') \cite{key-21}, in which experimental
data is naturally recorded, and the description in terms of reactant
numbers (``counting process'') \cite{key-22}, through which mechanisms
are most conveniently expressed. 

In Section \ref{sec:Renewal-processes-in} we apply this formalism
to study a reaction network corresponding to a single enzyme. In such
a network, reactant numbers are either zero or one, and a non-zero
value of one reactant number implies zero values of all others. We
explore the consequences of this ``fermionic'' character and find
that, irrespective of the complexity of the network, turnovers are
always statistically independent and identically distributed, or,
in other words, constitute a renewal process \cite{key-23}. A single-enzyme
network, then, cannot show molecular memory. 

In Section \ref{sec:Transients,-non-renewal-and} we consider a network
consisting of replicas of single-enzyme networks, corresponding to
oligomeric enzymes with independent and identical binding sites. The
absence of ``fermionic'' character in these networks permits turnovers
to be statistically dependent and allows them to show molecular memory.
The statistical dependence decreases with the number of turnovers
and vanishes asymptotically. We characterize this transient with fading
memory through the conditional distribution of consecutive turnovers,
which we relate to measures of the single-enzyme network. This analysis
explains the counter-intuitive appearance of memory in a process whose
elementary steps, recalling that the CME describes a Markov process,
are memoryless. 

In Section \ref{sec:Statistical-measures} we discuss new statistical
measures that, contrary to existing measures \cite{key-24,key-25,key-26},
do not assume the statistical independence of turnovers. Our measures,
then, can be applied uniformly over the entire duration of the catalytic
process, both in the transient state with memory and the steady state
in which memory vanishes. We provide an expression for the enzymatic
velocity in terms of turnover times that reduces to the classical
expression in the thermodynamic limit and elucidates how this limit
is reached. 

In Section \ref{sec:Comparison-with-data} we compare our theory with
the classic experiment on $\beta$-galactosidase \cite{key-12}, a
tetrameric enzyme, and find excellent agreement with four replicas
of a single-enzyme network with conformers and parallel pathways.
Saliently, we do not need to assume any ad-hoc distribution of reaction
rates \cite{key-27,key-28}: the ``dynamic disorder'' implied by
such a distribution is an emergent feature of our theory. 

We conclude, in section \ref{sec:Conclusion}, with a discussion on
how our theory can be extended to non-replica networks corresponding
to enzymes with interacting binding sites\textbf{. }
\begin{figure*}
\centering\includegraphics[clip,width=1\textwidth]{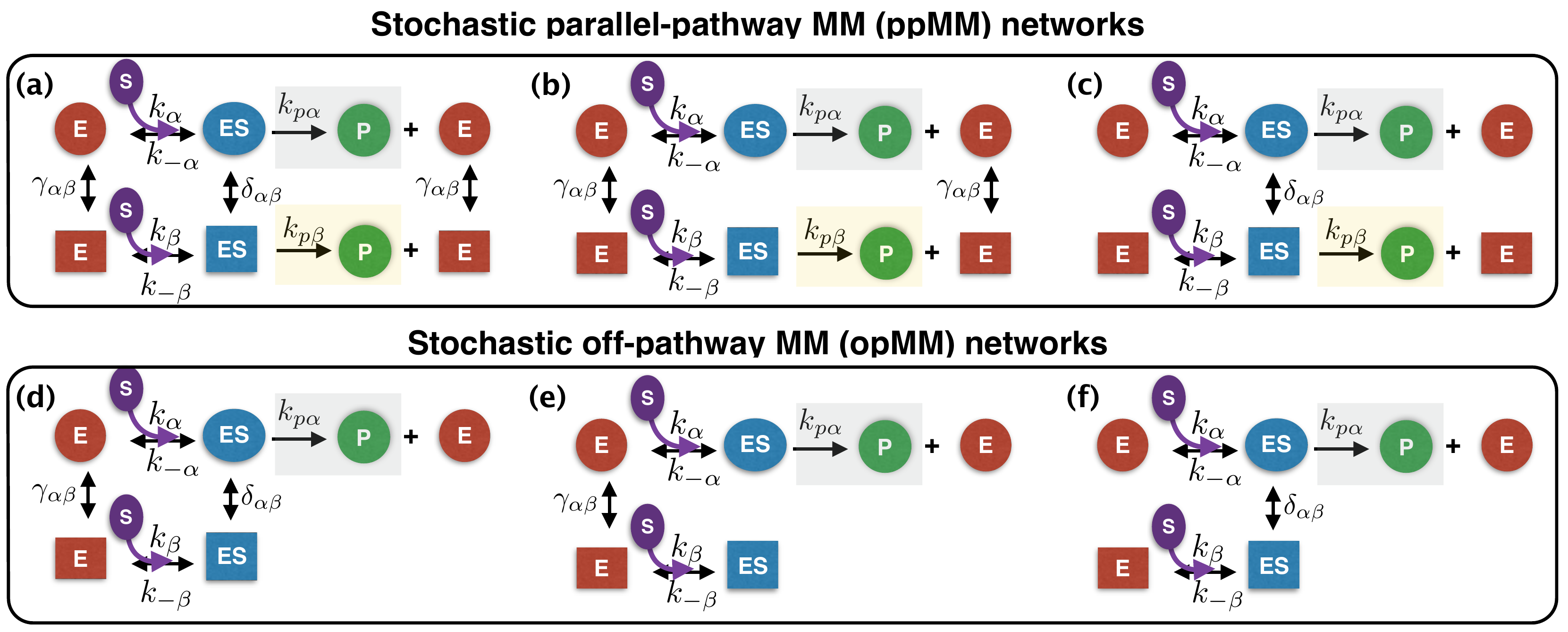}

\caption{Enzymatic networks for a single catalytic site, including conformational
fluctuations and parallel pathways to product formation. The general
network in (a) reduces to the special cases in (b) through (f) when
rate constants for the corresponding steps are set to zero. \label{fig:stochastic-networks}}
\end{figure*}

\section{Stochastic enzymatic networks \label{sec:Stochastic-enzymatic-networks}}

The stochastic description of chemical reactions begins with a set
of $k$ non-negative integers $\boldsymbol{n}=(n_{1},\ldots,n_{k})$
describing the number of molecules of the $k$-th species. Elementary
reactions

\begin{equation}
\sigma\text{-th reaction step}:\,\,\,\boldsymbol{n}\underset{t_{\sigma}^{\leftarrow}}{\stackrel{t_{\sigma}^{\rightarrow}}{\rightleftharpoons}}\boldsymbol{n}+\boldsymbol{r}_{\sigma}
\end{equation}
labelled by the index $\sigma$, take the state $\boldsymbol{n}$
to the state $\boldsymbol{n}+\boldsymbol{r}_{\sigma}$ where $\boldsymbol{r}_{\sigma}$
is a vector representing the integer changes of each species, as determined
by the reaction stoichiometry. The probability per unit time that
this reaction takes place is $t_{\sigma}^{\rightarrow}(\boldsymbol{n})$.
The corresponding backward reaction takes the state $\boldsymbol{n}+\boldsymbol{r}_{\sigma}$
to the state $\boldsymbol{n}$ at the rate $t_{\sigma}^{\leftarrow}(\boldsymbol{n}+\boldsymbol{r}_{\sigma})$.
The rates are combinatoric functions that follow from the law of mass
action. The probability $P(\boldsymbol{n},t)$ of being in the state
$\boldsymbol{n}$ at time $t$ is governed by the CME \cite{key-22,key-29,key-18}

\begin{align}
\partial_{t}P(\boldsymbol{n}) & =\sum_{\sigma}t_{\sigma}^{\rightarrow}(\boldsymbol{n}-\boldsymbol{r}_{\sigma})P(\boldsymbol{n}-\boldsymbol{r}_{\sigma})-t_{\sigma}^{\leftarrow}(\boldsymbol{n})P(\boldsymbol{n})\nonumber \\
 & +\sum_{\sigma}t_{\sigma}^{\leftarrow}(\boldsymbol{n}+\boldsymbol{r}_{\sigma})P(\boldsymbol{n}+\boldsymbol{r}_{\sigma})-t_{\sigma}^{\rightarrow}(\boldsymbol{n})P(\boldsymbol{n})
\end{align}
which is a system of coupled ordinary differential equations, equal
in number to the number of distinct states of the network. 

Here, we consider enzymatic networks that contain the Michaelis-Menten
mechanism $E_{i}+S\rightleftharpoons ES_{i}\rightarrow E_{i}+P$ as
a basic motif while allowing for conformers $i=\alpha,\beta,\ldots$
and parallel pathways to product formation \cite{key-12,key-13}.
The state is described by the vector $\boldsymbol{n}=(n_{E_{\alpha}},n_{ES_{\alpha}},n_{E_{\beta}},n_{ES_{\beta}},\ldots,n)$
of non-negative integers comprising of, in obvious notation, the numbers
of enzyme and complex, of each conformational type, and of product.
Examples of such networks for the simplest case of two conformers
are shown in Fig.(\ref{fig:stochastic-networks}) for both parallel
and off-pathway kinetics \cite{key-30}. The corresponding rates are
listed in Table. (\ref{tab:PPMM-steps}). The bimolecular complexation
steps are replaced by pseudo-unimolecular steps with effective rate
constants denoted by primes. All rates are then linear in the state
vector $\boldsymbol{n}$. It is important to note that the rates do
not depend on the number of products. 
\begin{table}
\centering%
\begin{tabular}{|c|c|c|c|}
\hline 
Step & $\boldsymbol{r}_{\sigma}$ & $t_{\sigma}^{\rightarrow}(\boldsymbol{n})$ & $t_{\sigma}^{\leftarrow}(\boldsymbol{n})$\tabularnewline
\hline 
\hline 
$E_{\alpha}\rightleftharpoons E{}_{\beta}$ & $(-1,0,1,0,0)$ & $\gamma_{\alpha\beta}n_{E_{\alpha}}$ & $\gamma_{\alpha\beta}n_{E_{\beta}}$\tabularnewline
\hline 
$ES_{\alpha}\rightleftharpoons ES_{\beta}$ & $(0,-1,0,1,0)$ & $\delta_{\alpha\beta}n_{ES_{\alpha}}$ & $\delta_{\alpha\beta}n_{ES_{\beta}}$\tabularnewline
\hline 
$E_{\alpha}\rightleftharpoons ES_{\alpha}$ & $(-1,1,0,0,0)$ & $k_{\alpha}^{\prime}n_{E_{\alpha}}$ & $k_{-\alpha}n_{ES_{\alpha}}$\tabularnewline
\hline 
$E_{\beta}\rightleftharpoons ES_{\beta}$ & $(0,0,-1,1,0)$ & $k_{\beta}^{\prime}n_{E_{\beta}}$ & $k_{-\beta}n_{ES_{\beta}}$\tabularnewline
\hline 
$ES_{\alpha}\rightarrow P+E{}_{\alpha}$ & $(1,-1,0,0,1)$ & $k_{p\alpha}n_{ES_{\alpha}}$ & $0$\tabularnewline
\hline 
$ES_{\beta}\rightarrow P+E{}_{\beta}$ & $(0,0,1,-1,1)$ & $k_{p\beta}n_{ES_{\beta}}$ & $0$\tabularnewline
\hline 
\end{tabular}\caption{Elementary reaction steps and their rates for the single-site network
labelled (a) in Fig. (\ref{fig:stochastic-networks}). The networks
labelled (b) through (c) are obtained by setting corresponding rate
parameters to zero. The forward reaction takes the state $\boldsymbol{n}=(n_{E_{\alpha}},n_{ES_{\alpha}},n_{E_{\beta}},n_{ES_{\beta}},n)$
to the state $\boldsymbol{n}+\boldsymbol{r}_{\sigma}$. The pseudo-first-order
rate constants for the forward reaction are $k_{\alpha}^{\prime}=k_{\alpha}[S]$
and $k_{\beta}^{\prime}=k_{\beta}[S]$. All rates are independent
of the number of products $n$. }
\label{tab:PPMM-steps}
\end{table}

It is convenient to partition the state vector into $\boldsymbol{n}=(\boldsymbol{n}^{\star},n)$,
where $\boldsymbol{n}^{\star}=(n_{E_{\alpha}},n_{ES_{\alpha}},n_{E_{\beta}},n_{ES_{\beta}},\ldots)$
are ``hidden'' state components unobserved in experiment, and $n$
is the ``observed'' product state visible through fluorescence bursts.
The hidden state vector has $2l$ components in a network with $l$
conformers. For a network with $\nu$ enzymes, or one oligomeric enzyme
with $\nu$ active sites, mass conservation implies that the sum of
the number of enzymes in the uncomplexed and complexed states must
sum to $\nu$: $n_{E_{\alpha}}+n_{E_{\beta}}+\ldots+n_{ES_{\alpha}}+n_{ES_{\beta}}+\ldots=\nu$.
For a single enzyme (or active site), this implies that $n_{E_{\alpha}}+n_{E_{\beta}}+\ldots+n_{ES_{\alpha}}+n_{ES_{\beta}}+\ldots=1$.
Therefore, the components of the hidden state vector in a single-enzyme
network have a ``fermionic'' character, where the components only
take the values zero or one and only one component is non-zero at
any point in time. Mass conservation also implies that the number
of hidden states is finite and equal to the number of compositions
of $\nu$ into $l$ parts. For a single enzyme with $l$ conformers,
this gives $2l$ states. We shall return to this important property
below. 

Unlike the hidden components, the number of products $n$ can take
values from zero to infinity. The probability of their being $n$
products at time $t$ is obtained by marginalizing the reactant probability
over the hidden states, 

\begin{equation}
P(n,t)=\sum_{\boldsymbol{n}^{\star}}P(\boldsymbol{n},t).
\end{equation}
This is the fundamental probability distribution in the counting process
description of turnovers. The expectation with respect to this probability
distribution of the mean and variance of $n$ define the enzymatic
velocity and Fano factor \cite{key-31},

\begin{equation}
V(t)=\frac{d}{dt}\langle n\rangle,\quad\rho(t)=\frac{\langle n^{2}\rangle-\langle n\rangle^{2}}{\langle n\rangle}\quad(t\geq0)\label{eq:measues-cp}
\end{equation}
Such quantities have been calculated for a variety of networks beginning
with the work of Bartholomay. However, as mentioned in the Introduction,
they are not directly relevant to single-enzyme experiments which
record the times $T_{p}$ at which turnovers occur, rather than the
number of turnovers at time $t$. Here $p=1,2,\cdots$ is the turnover
number index. This motivates the study of the point process of turnovers,
for which we now introduce the fundamental probability distributions
\cite{key-21}. 

We define the turnover time for the $p$-th product, $T_{p},$ to
be the smallest value of $t$ such that $n\geq p$, or more precisely,
$T_{p}=\text{inf}\{t>0:n(t)\geq p\}$. The cumulative distribution
of the $p$-th turnover time $T_{p}$ is denoted by $P(T_{p}\leq t)$.
This defines the survival probability $P(T_{p}>t)=1-P(T_{p}\leq t)$
and probability density $w_{T_{p}}(t)dt=P(t<T_{p}\leq t+dt)$. The
expectation of $T_{p}$ with respect to the probability density defines
the mean turnover time and the randomness parameter for the $p$-th
turnover,

\begin{equation}
\mu_{p}=\langle T_{p}\rangle,\quad r_{p}=p\frac{\langle T_{p}^{2}\rangle-\langle T_{p}\rangle^{2}}{\langle T_{p}\rangle^{2}}\quad(p=1,2,\ldots)\label{eq:measures-pp}
\end{equation}
While the current definition of the randomness parameter is $p$ independent\textbf{
$r=\langle\tau^{2}\rangle-\langle\tau\rangle^{2}/\langle\tau\rangle^{2}$}
\cite{key-32,key-33}, the factor of $p$ in the above definition
of the randomness parameter is introduced for reasons that will become
apparent in Section \ref{sec:Statistical-measures}. Higher moments
can be studied but, to the best of our knowledge, have not been measured
in experiment. 

To quantify statistical dependences, it is convenient to define the
waiting time between turnovers,
\begin{equation}
\tau_{p}=T_{p}-T_{p-1},\label{eq:pth-wating time}
\end{equation}
and study their joint density distributions, $w(\tau_{1},\tau_{2},\ldots)$.
The marginal distributions $w(\tau_{p})$ describe the statistics
of individual turnovers while the joint distributions $w(\tau_{p},\tau_{q})$
describe the statistics of pairs of turnovers. It is convenient to
write this joint distribution as 
\begin{equation}
w(\tau_{p},\tau_{q})=w(\tau_{p})g(\tau_{p},\tau_{q})w(\tau_{q})\label{eq:joint-wtd}
\end{equation}
 so that statistical dependences are contained in $g(\tau_{p},\tau_{q})$.
Pairs of turnovers are statistically independent if and only if $g(\tau_{p},\tau_{q})=1$
for all $p$ and $q$. The correlation function
\begin{equation}
C_{pq}=\langle\tau_{p}\tau_{q}\rangle=c(\tau_{p},\tau_{q})\langle\tau_{p}\rangle\langle\tau_{q}\rangle\label{eq:corrpq}
\end{equation}
serves as a second-order statistic for identifying statistical dependences.
The expectations are with respect to the joint distribution $w(\tau_{p},\tau_{q}).$
Statistical dependences and molecular memory imply $c(\tau_{p},\tau_{q})\ne1$. 

The question naturally arises as to how the probability distributions
for the counting process, $P(n,t)$, and the point process, $P(T_{p}\le t)$,
together with their summary statistics, are related to each other
and to the underlying CME which is the generative process that underlies
both distributions. We provide the answer below. 

From the definitions of the random variables $n$ and $T_{p}$, it
is clear that at any time $t$,
\begin{equation}
T_{p}\le t\Longleftrightarrow n(t)\ge p
\end{equation}
or, in other words, these two events are equal in probability. Since
the event $n(t)<p$ is their complement, we have 

\begin{equation}
P(T_{p}\leq t)=P(n\geq p,t)=1-P(n<p,t).
\end{equation}
Since the product states are mutually exclusive, we have $P(n<p,t)=\sum_{n=0}^{p-1}P(n,t)$.
Combining this with the marginal expression for $P(n,t)$ we obtain
\begin{equation}
P(T_{p}\leq t)=1-\sum_{n=0}^{p-1}\sum_{\boldsymbol{n}^{\star}}P(\boldsymbol{n},t).\label{eq:central-result}
\end{equation}
This relation between the turnover time distribution and the solution
of the CME is the central result of this section. It is applicable
to networks of arbitrary complexity and provides the sought after
connection between the statistics of turnovers and reaction mechanisms.
The probability density follows upon differentiation,

\begin{equation}
w_{T_{p}}(t)=-\sum_{n=0}^{p-1}\sum_{\boldsymbol{n}^{\star}}\partial_{t}P(\boldsymbol{n},t),
\end{equation}
and is often more convenient for comparison with experimental data,
when the latter is presented in the form of a probability density.
A special case of this relation was first obtained in \cite{key-15}. 

We have not been able to find a relation of this generality that relates
the waiting time distributions $w(\tau_{p})$ and $w(\tau_{p},\tau_{q})$
to the solution of the CME. However, in particular instances, where
the network is ``fermionic'' or has a ``replica'' character, relations
to the underlying CME can be found, as we show in Sections \ref{sec:Renewal-processes-in}
and \ref{sec:Transients,-non-renewal-and}, respectively. 

\section{Renewal statistics in single-enzyme networks \label{sec:Renewal-processes-in}}

As we noted above, hidden states in a single-enzyme network have a
``fermionic'' character: the components of the hidden state vector
can only take the values zero or one, and only one component can be
non-zero at any time. This implies that immediately after the conclusion
of a turnover, say the $p$-th, the network is in a state corresponding
to a single uncomplexed enzyme. To elaborate, consider the MM network
with state vector $\boldsymbol{n}=(n_{E},n_{ES},n)$ and hidden state
vector $\boldsymbol{n}^{\star}=(n_{E},n_{ES})$. The two allowed hidden
states are $(1,0)$ and $(0,1)$ corresponding to uncomplexed and
complexed enzyme. Labeling these by $E$ and $ES$, the allowed states
of the network are $(E,n)$ and $(ES,n)$. At the conclusion of the
$p$-th turnover at $t=T_{p}$ the network is in the state $(E,p)$
and so, taking limits from above, 
\begin{equation}
\lim_{t\rightarrow T_{p}^{+}}P(E,p,t)=1.
\end{equation}
 For the two-conformer ppMM network with state vector $\boldsymbol{n}=(n_{E_{\alpha}},n_{ES_{\alpha}},n_{E_{\beta}},n_{ES_{\beta}},n)$
and hidden state vector $\boldsymbol{n}^{\star}=(n_{E_{\alpha}},n_{ES_{\alpha}},n_{E_{\beta}},n_{ES_{\beta}})$
the four allowed hidden states are $(1,0,0,0)$, $(0,1,0,0),$ $(0,0,1,0)$
and $(0,0,0,1)$. Labeling thse by $E_{\alpha},$ $ES_{\alpha}$,
$E_{\beta}$ and $ES_{\beta}$ , the allowed states of the network
are $(E_{\alpha},n)$, $(ES_{\alpha},n),$ $(E_{\beta},n)$ and $(ES_{\beta},n)$.
At the conclusion of the $p$-th turnover, the network is either in
the state $(E_{\alpha},p)$ or in the state $(E_{\beta},p)$ and so
\begin{equation}
\lim_{t\rightarrow T_{p}^{+}}P(E_{\alpha},p,t)+P(E_{\beta},p,t)=1
\end{equation}
More generally, for any single-enzyme network the states can be labelled
by the conformation of the enzyme and the number of products and at
the conclusion of a turnover, the network is surely in one of the
uncomplexed states with the total probability is partitioned between
those states. As we show below, this recurrent return to a fixed subset
of the hidden states, together with the structure of the CME for such
networks, implies that conditioning on a turnover makes the future
independent of the past. This results in turnovers that are statistically
independent, with waiting time distributions that are identically
distributed. Single-enzyme turnovers, therefore, form a renewal process
and cannot show memory \cite{key-23}. 
\begin{figure*}
\centering\includegraphics[clip,width=1\textwidth]{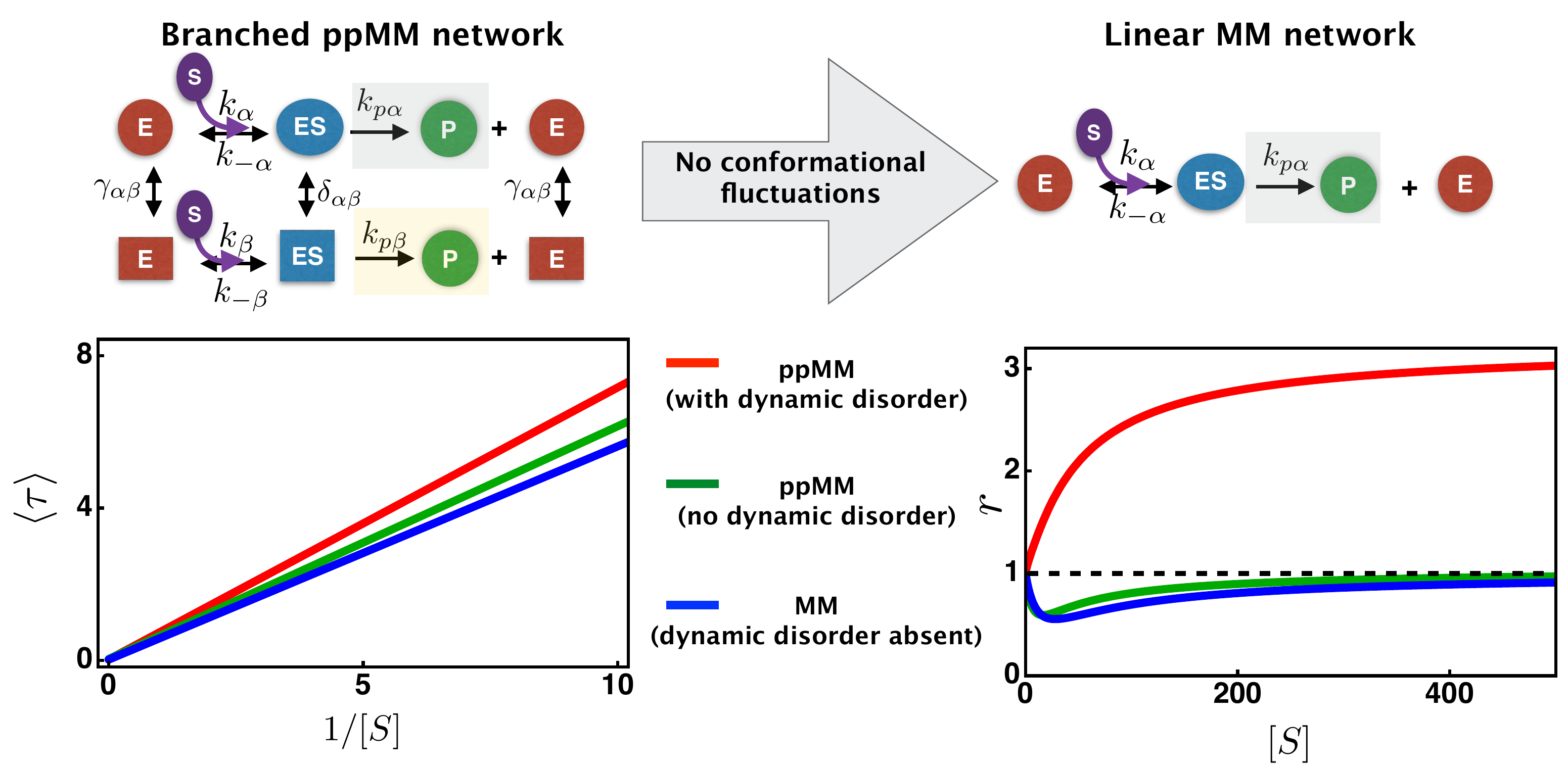}

\caption{Distinction between the summary statistics of branched and single
pathway stochastic networks as a function of substrate concentration.
Top panel shows how a parallel-pathway MM (ppMM) network reduces to
a single-pathway MM network when conformational fluctuations are disallowed.
Bottom left panel shows the single-enzyme Lineweaver-Burk plot for
the variation of inverse single-enzyme velocity, $\langle\tau\rangle$,
with $1/[S]$. The plot is linear, and thus the variation of single-enzyme
velocity with $[S]$ is hyperbolic, irrespective of the network complexity.
Bottom right panel shows how the variation of the randomness parameter
$r$ with $[S]$ provides a quantitative measure to discern single-enzyme
network topologies. For the ppMM network, the rate constants conditions
that favor parallel-pathway for product formation can yield $r\geq1$
(dynamic disorder). The rate constant conditions that disallow the
latter always yield $r\leq1$. For a linear MM network, irrespective
of the rate parameter conditions, there is no dynamic disorder as
$r$ does not exceed one.\label{fig:single-and-replica-networks}}
\end{figure*}

In the absence of memory, attention can be focussed entirely on the
statistics of the waiting time, which, since it is identically distributed
for all $p,$ we simply denote by $\tau$. We summarize our two main
results before providing explicit results for the MM and ppMM networks.
First, we show that for a model with $l$ conformers the waiting time
distribution is a sum of $2l$ exponentials whose time constants are
eigenvalues of a $2l\times2l$ matrix related to the CME. The behavior
of the waiting time distribution is related to the spacing of these
eigenvalues. For well-separated eigenvalues, the exponential with
the lowest time constant is dominant, but for closely spaced eigenvalues
all the $2l$ exponentials contribute. This multi-exponentiality leads
to variances that are large compared to the squared mean and, hence,
to a randomness parameter that exceeds unity. Our analysis thus transparently
relates ``dynamic disorder'' to reaction mechanisms with fixed rate
constants, in contrast to fluctuating rate parameters \cite{key-27,key-28}.
Second, we show that the randomness parameter for $\tau$ is a sensitive
measure of network topology. It is known that a Markov chain comprising
of a linear network of arbitrary complexity always yields $r\leq1$,
bounded below by the inverse of the number of rate determining steps
$\sigma$: $r\geq\frac{1}{\sigma}$ \cite{key-34}. The minimum is
attained for a linear sequence of states with equal rate constants,
first studied by Erlang \cite{key-35}. For a single step reaction
or a linear network with a single rate determining step, thus, $r=1$.
We find that a branched topology is necessary, but not sufficient,
to obtain $r>1$. Our explicit calculation for the ppMM network shows
that $r$ can vary continuously from $r<1$ to $r>1$ as the substrate
concentration is increased. Thus, networks can be rationally designed
to yield a desired value of the randomness parameter. 

Consider, now, the CME for the MM network, written in terms of the
labels $E$,$ES$, and $n$: \begin{widetext}

\begin{align}
\partial_{t}P(E,n) & =-k_{a}P(E,n)+k_{-1}P(ES,n)+k_{2}P(ES,n-1)\nonumber \\
\partial_{t}P(ES,n) & =+k_{a}P(E,n)-k_{-1}P(ES,n)-k_{2}P(ES,n),\quad\quad n=0,1,2,\ldots
\end{align}
\end{widetext}It is understood that states with $n<0$ have zero
probability. This is an infinite system of autonomous linear differential
equations whose solution can be obtained using the technique of generating
functions. A great simplification results when we recognize the following
two features. First, conditioning the system on the $p-$th turnover
at $t=T_{p}$ collapses the probability on the state $(E,p)$ so that
$P(E,p)=1$ and all other probabilities are zero. Since probabilities
only flow into states with increasing number of products, this implies
that probabilities of all states with $n<p$ remain zero subsequently.
The future is made conditionally independent of the past. Second,
the distribution of the $(p+1)$-th turnover $T_{p+1}$, conditioned
on the $p-$th turnover at $T_{p}$, is governed by the same set of
equations and initial conditions as the first turnover $T_{1}$, conditioned
on the initial state at $t=0$. This implies that $T_{1}$ and $T_{p+1}-T_{p}$
are equal in distribution for all $p$. Therefore, the waiting times
$\tau_{p}$ are independent and distributed identically to $T_{1}$. 

From Eq. (\ref{eq:central-result}) the cumulative distribution of
$T_{1}$ is 
\begin{equation}
P(T_{1}<t)=1-[P(E,0,t)+P(ES,0,t)]
\end{equation}
and from the CME these two probabilities obey 
\[
\left[\begin{array}{c}
\partial_{t}P(E,0,t)\\
\partial_{t}P(ES,,0,t)
\end{array}\right]=\left[\begin{array}{cc}
-k_{1}^{\prime} & k_{-1}\\
k_{1}^{\prime} & -(k_{-1}+k_{2})
\end{array}\right]\left[\begin{array}{c}
P(E,0,t)\\
P(ES,0,t)
\end{array}\right]
\]
with initial condition $P(E,0)=1$ and $P(ES,0)=0$ at $t=0.$ This
is a system of ordinary differential equations for the vector $\boldsymbol{P}(t)=[P_{E}(t),P_{ES}(t)]$
, where $P(E,0,t)$ is abbreviated as $P_{E}(t)$ \emph{etc, }with
system matrix 
\[
\boldsymbol{L}=\exp\left[\begin{array}{cc}
-k_{1}^{\prime}t & k_{-1}t\\
k_{1}^{\prime}t & -(k_{-1}+k_{2})t
\end{array}\right].
\]
The solution is obtained in terms of the matrix exponential as $\boldsymbol{P}(t)=\exp(\boldsymbol{L}t)\cdot\boldsymbol{P}(0)$
with the explicit result, 
\begin{align*}
P_{E}(t)= & \frac{1}{2A}\left[(A+B-C)e^{-(B+A)t}+(A-B+C)e^{-(B-A)t}\right]\\
P_{ES}(t)= & \frac{k_{1}^{\prime}}{2A}\left[e^{-(B-A)t}-e^{-(B+A)t}\right],
\end{align*}
where $k_{1}^{\prime}=k_{1}[S]$, $2A=\sqrt{(k_{1}^{\prime}+k_{-1}+k_{2})^{2}-4k_{1}^{\prime}k_{2}}$
and $2B=\left[k_{1}^{\prime}+k_{-1}+k_{2}\right]$, and $C=k_{-1}+k_{2}$.
From this it is clear that the cumulative distribution is a sum of
two exponentials whose time constants are determined by the eigenvalues
of the $2\times2$ matrix in the argument of the matrix exponential.
The waiting time distribution follows on differentiating the cumulative
distribution, 

\begin{equation}
w(\tau)=\frac{k_{2}k_{1}^{\prime}}{2A}\left[e^{(A-B)\tau_{1}}-e^{-(A+B)\tau}\right]\label{eq:single-site wtd}
\end{equation}
and the corresponding mean and randomness parameter are 

\begin{gather}
\langle\tau\rangle=\frac{1}{k_{2}}+\frac{k_{-1}+k_{2}}{k_{1}}\frac{1}{[S]},\label{eq:single-site mwt}\\
r=\frac{(k_{1}[S]+k_{-1})^{2}+2k_{2}k_{-1}+k_{2}^{2}}{(k_{1}[S]+k_{-1}+k_{2})^{2}}.\label{eq:single-site rp}
\end{gather}
The variation of these is shown in Fig. (\ref{fig:single-and-replica-networks})
as a function of substrate concentration. The mean waiting time has
a hyperbolic dependent on the substrate concentration, of the Michaelis-Menten
form, and the randomness parameter is always less the unity, in agreement
with previous analysis \cite{key-32,key-33}. 

How, now, are these results altered when the network topology is altered
to allow for conformational fluctuations? The master equation for
the general network shown in Fig. (\ref{fig:stochastic-networks}a)
is 

\begin{widetext}

\begin{alignat}{1}
\partial_{t}P(E_{\alpha},n) & =k_{-\alpha}P(ES_{\alpha},n)+\gamma_{\alpha\beta}P(E_{\beta},n)-(k_{\alpha}^{\prime}+\gamma_{\alpha\beta})P(E_{\alpha},n)+k_{p\alpha}P(ES_{\alpha},n-1)\nonumber \\
\partial_{t}P(E_{\beta},n) & =k_{-\beta}P(ES_{\beta},n)+\gamma_{\alpha\beta}P(E_{\alpha},n)-(k_{\beta}^{\prime}+\gamma_{\alpha\beta})P(E_{\beta},n)+k_{p\beta}P(ES_{\beta},n-1)\nonumber \\
\partial_{t}P(ES_{\alpha},n) & =k_{\alpha}^{\prime}P(E_{\alpha},n)+\delta_{\alpha\beta}P(ES_{\beta},n)-(k_{-\alpha}+\delta_{\alpha\beta}+k_{p\alpha})P(ES_{\alpha},n)\nonumber \\
\partial_{t}P(ES_{\beta},n) & =k_{\beta}^{\prime}P(E_{\beta},n)+\delta_{\alpha\beta}P(ES_{\alpha},n)-(k_{-\beta}+\delta_{\alpha\beta}+k_{p\beta})P(ES_{\beta},n),\quad\quad n=0,1,2,\ldots
\end{alignat}
\end{widetext}Conditioning on a turnover, as before, reduces the
CME to a system for four coupled differential equations for the components
of the vector 
\[
\boldsymbol{P}(t)=[P_{E_{\alpha}}(t),P_{ES_{\alpha}}(t),P_{E_{\beta}}(t),P_{ES_{\beta}}(t)]
\]
in the abbreviated notation introduced above. The solution is given
in term of the exponential of the $4\times4$ matrix system matrix
so that the cumulative distribution is a sum of four exponential terms.
The expression for the waiting time distribution is obtained by differentiation
as before. The expressions, being unwieldy, are provided in the Supplementary
Information (SI). The results are shown in Fig. (\ref{fig:single-and-replica-networks})
for the general network for two sets of rate constants. The mean continues
to have a hyperbolic dependence on the substrate concentration but
now the randomness parameter can yield values that are lesser or greater
than unity, depending on the choice of rate constants. We find that
rate constants that tend to suppress parallel pathways, \emph{i.e.
}to make the system matrix block diagonal, correspond to randomness
parameters less than unity. In this limit, there is little to distinguish
between linear and branched topologies. On the other hand, for rate
constants that promote parallel pathways, \emph{i.e. }to make the
system matrix dense, correspond to randomness parameters greater than
unity. This is the regime of dynamic disorder and our results show
that such effects can be obtained without imputing any ad-hoc fluctuations
on the rate constants themselves, but by simply allowing for a change
in network topology. 

The conditional independence of turnovers, due to the ``fermionic''
nature of the states, implies that single-enzyme networks can never
show molecular memory. We now turn to networks in which the ``fermionic''
nature is lost, in the simplest possible way, by considering replicas
of single-enzyme networks.%
\emph{}%

\section{replica networks\label{sec:Transients,-non-renewal-and}}

\begin{figure}
\includegraphics[scale=0.32]{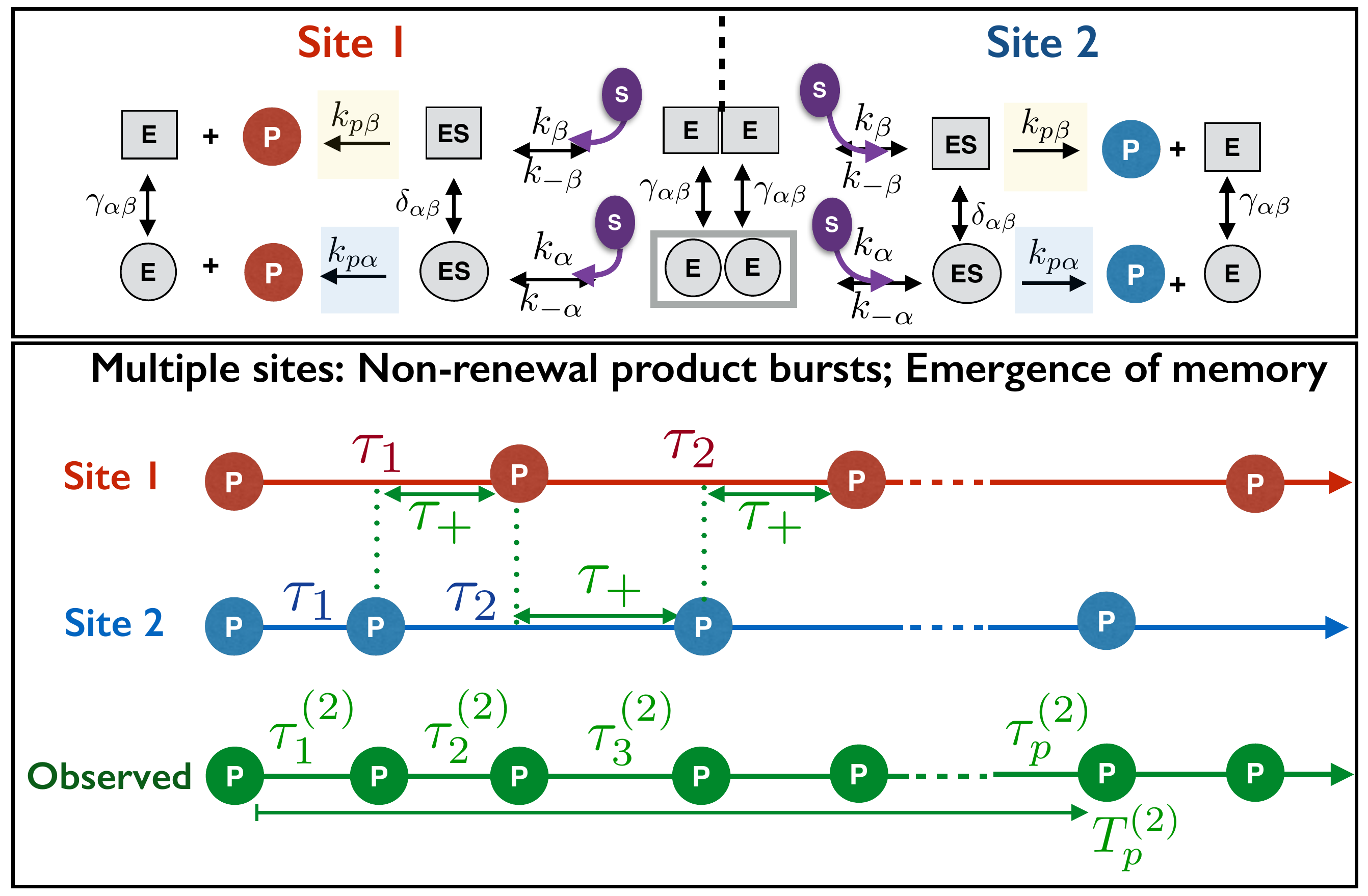}

\caption{The sequence of turnovers for a pair of active sites that process
substrates in parallel. The waiting times are denoted by $\tau$ and
the forward recurrence times (see text) by $\tau_{+}$. The observed
turnovers cannot distinguish which of the two active sites yielded
the product and, hence, the turnover process is a pooling of the two
independent turnover processes for each site. The generalization to
more than two sites is obvious. \label{fig:pooled-process}}
\end{figure}

Consider now a pair of sites on a single enzyme, as shown in Fig.(\ref{fig:pooled-process})
in red and blue, each governed by identical ppMM mechanisms and catalysing
substrates independently. It is neither possible, nor relevant, to
distinguish such products by their site of production and the \emph{observed}
process of turnover, shown in green, is a ``pooling'' of the independent
turnover processes at each site. The total number of products in the
pooled process at time $t$ is the sum of the number of products at
each site. Since the latter are independent random variables, the
statistics of their sum can be simply obtained from the individual
statistics. Therefore, the counting process of pooled turnovers is
simple. However, the point process of pooled turnovers has a less
simple relation to the individual point processes, as we explain below.

Returning to Fig.(\ref{fig:pooled-process}), assume that both sites
start from identical initial conditions of being in uncomplexed states
and denote by $\tau_{p}$ the $p$-th waiting time at any one of the
identical sites and $\tau_{p}^{(2)}$ the $p$-th waiting time of
the pooled process, superscript indicating that a pair of processes
are pooled. Then, the waiting time $\tau_{1}^{(2)}$ for the first
product of the pooled process is the shorter of the first waiting
times $\tau_{1}$ at each of the sites. For the second and subsequent
turnovers, it is necessary to introduce the notion of the \emph{forward
recurrence time $\tau_{+}$ , }that is the waiting time to the \emph{next}
product starting at an arbitrary time $t$. The distribution of $\tau_{+}$,
$P(s<\tau_{+}<s+ds|t)\equiv w_{+}(s|t)ds$, is conditional on the
time $t$ and this conditional dependence is crucial in what follows.
In terms of the forward recurrence time, the waiting time $\tau_{2}^{(2)}$
of the second product is the shorter of the waiting time $\tau_{2}$
at the site that produced the first product and the\emph{ }forward
recurrence time\emph{ $\tau_{+}$} of the site that did not. For the
example in Fig.(\ref{fig:pooled-process}), these are the first and
second sites, respectively. Generalizing, the waiting time $\tau_{p}$
of the $p$-th product is the shorter of the waiting at one site and
the forward recurrence time at the other site, where the recurrence
time is measured from the last turnover at $t=T_{p-1}^{(2)}$. Since
a waiting time $\tau_{p}^{(2)}$ exceeding $s$ implies that\textbf{\emph{
}}\emph{both }the waiting time $\tau_{p}$ and the recurrence time
$\tau_{+}$ measured from the last turnover exceed $s,$ that is
\begin{equation}
\tau_{p}^{(2)}>s\Longleftrightarrow(\tau_{p}>s)\,\,\text{AND\,\,(\ensuremath{\tau_{+}>s)} }
\end{equation}
we immediately obtain for the survival probability of $\tau_{p}^{(2)}$
the relation

\begin{equation}
P(\tau_{p}^{(2)}>s|T_{p-1}^{(2)})=P(\tau>s)P(\tau_{+}>s|T_{p-1}^{(2)})
\end{equation}
This basic result shows that for a pooled process, the $p$-the waiting
time is, in general, conditionally dependent on the time at which
the $(p-1)$-th turnover takes place. Therefore, the very act of pooling
provides a mechanism by which a future waiting time can become conditionally
dependent on past waiting times, or, in other words, for the emergence
of molecular memory. 

It is desirable, if possible, to relate this conditional dependence
to properties of the renewal process at each site. We now show that
this is, indeed, possible. Defining the waiting time distribution
of the pooled process as $P(s<\tau_{p}^{(2)}<s+ds|T_{p-1}^{(2)})\equiv w_{\pi}(s|T_{p-1}^{(2)})ds$
and differentiating both sides of the above we obtain 

\begin{eqnarray}
w_{\pi}(s|T_{p-1}^{(2)}) & = & w(s)[1-P(\tau_{+}<s|T_{p-1}^{(2)})]+\nonumber \\
 &  & [1-\int_{0}^{s}w(s')ds']w_{+}(s|T_{p-1}^{(2)})\label{eq:memory-2replica}
\end{eqnarray}

In the above, the distribution of waiting times $w(s)$ is known from
the analysis of the previous section and it only remains to determine
the distribution $w_{+}(s|t)$ of the recurrence time. For a renewal
process, the distribution of the forward recurrence time is related
to the waiting time distribution and the enzymatic velocity by \cite{key-23}
\begin{equation}
w_{+}(s|t)=w(t+s)+\int_{0}^{t}V(t-s^{\prime})w(s+s^{\prime})ds^{\prime}.\label{eq:rectimedist}
\end{equation}
and this determines the waiting time distribution of the pooled process
completely in terms of quantities that can be calculated from the
component renewal processes. 

How does the conditional dependence of the waiting time vary with
the number of turnovers ? Since the entire conditional dependence
derives from the distribution of the recurrence time, it is sufficient
to examine its conditional dependence. For large times $t\gg T^{\star}$,
it is known from both the deterministic and stochastic analysis that
the enzymatic velocity becomes a constant, that is $\lim_{t\rightarrow\infty}V(t)=V_{ss}$.
As a consequence,  $\lim_{t\rightarrow\infty}w_{+}(s|t)=V_{ss}[1-W(s)]$
is independent of $t.$ In this limit, each site is in ``equilibrium'',
and the memory of the initial state of the process, which began with
all sites free, is erased . Thus, the conditioning of the recurrence
time of one site by the turnover time of another site, together with
the deterministic initial condition, provides a mechanism for the
statistical dependence between waiting times in the pooled process
and of the emergence of molecular memory. 

Extending this argument to $\nu$ binding sites, and denoting pooled
quantities with the superscript $\nu$, it is clear that the waiting
time $\tau_{p}^{(\nu)}$ for the $p$-th product is the shortest of
a single waiting time and $\nu-1$ recurrence times conditioned on
$t=T_{p-1}^{(\nu)}$ . Since $\tau_{p}^{(\nu)}$ being longer than
$s$ implies both the waiting time \emph{and} the $\nu-1$ recurrence
times are longer than $s$, we have

\begin{equation}
P(\tau_{p}^{(\nu)}>s|T_{p-1}^{(\nu)})=P(\tau>s)[P(\tau_{+}>s|T_{p-1}^{(\nu)})]^{\nu-1}\label{eq:memory-formula-1}
\end{equation}
where the factorizations on the right follow from independence and
identity of the $\nu$ binding sites. Defining the waiting time distribution
of the pooled process as $P(s<\tau_{p}^{(\nu)}<s+ds|T_{p-1}^{(\nu)})\equiv w_{\pi}(s|T_{p-1}^{(\nu)})ds$
and differentiating both sides of the above yields an explicit expression
for $w_{\pi}(s|T_{p-1}^{(\nu)})$ in terms of the key single-site
measures, the enzymatic velocity $V(t)$, the waiting time distribution
$w(\tau)$, and the recurrence time distribution $w_{+}(s|t)$,

\begin{widetext}

\begin{equation}
w_{\pi}(s|T_{p-1}^{(\nu)})=w(s)[1-P(\tau_{+}<s|T_{p-1}^{(\nu)})]^{\nu-1}+(\nu-1)[1-\int_{0}^{s}w(s')ds'][1-P(\tau_{+}<s|T_{p-1}^{(\nu)})]^{\nu-2}w_{+}(s|T_{p-1}^{(\nu)}).
\end{equation}
 \end{widetext}. 

In this expression, $\tau_{p}^{(\nu)}$ is conditionally dependent
on $T_{p-1}^{(\nu)},$ and therefore on the previous waiting times
$\tau_{1}^{(\nu)}\ldots\tau_{p-1}^{(\nu)}$, as long as $w_{+}(s|t)$
is dependent on $t$. Thus, the emergence of memory can now be traced
explicitly to the transient in the enzymatic velocity, starting from
the deterministic initial condition. Evaluating $w_{\pi}(s|T_{p-1}^{(\nu)})$
numerically for the MM mechanism for $T_{p-1}^{(\nu)}$ in the transient,
crossover, and steady-state regimes quantitatively confirms this qualitative
picture, Fig. (\ref{fig:Heat-maps}) in the SI. 

The waiting times become independent and identically distributed when
the enzymatic velocity at each site reaches the steady state value.
Then, inserting asymptotic form of the recurrence time distribution,
we obtain $w_{\pi}^{(\nu)}(s)=-\frac{d}{ds}\left[P(\tau>s)\left\{ V_{ss}\int_{s}^{\infty}P(\tau>s')ds\right\} ^{\nu-1}\right]$
giving the waiting time distribution in terms of survival probability
and the steady-state enzymatic velocity at each site. This is the
enzymatic analog of a well-known result in renewal theory \cite{key-36}. 

\section{Statistical measures\label{sec:Statistical-measures}}

In classical deterministic enzyme kinetics, the enzymatic velocity
of $N$ independent and identical enzymes, $V(t)=d_{t}\langle n\rangle$,
is a statistical measure of mean rate of product formation. The approach
to steady-state is then marked by the asymptotic limit, $V_{ss}=\lim_{t\rightarrow\infty}V(t)$,
in which the enzymatic velocity reaches its equilibrium value and
becomes time independent. For $N\gg1$, this asymptotic limit is realized
at the onset of the reaction \cite{key-1}. This implies that the
initial mean rate of product formation, \emph{i.e.} the counting process
alone, is sufficient to yield the steady-state enzymatic velocity
$V_{ss}=\text{\ensuremath{\left.d_{t}\langle n\rangle\right|_{t\rightarrow0}}}$
, and the transient regime remains unobserved \cite{key-17}. 

In stochastic enzyme kinetics, in contrast, the statistical measures
of counting and point processes for means and fluctuations, introduced
in Section \ref{sec:Stochastic-enzymatic-networks}, seem to provide
an alternative description of product turnover kinetics in the number
and time domain, respectively. It is pertinent to ask, then, how these
seemingly unrelated statistical measures can be formally linked to
demarcate the transient and steady-state regimes in enzyme kinetics
at the molecular level, and how these results can be reconciled with
the classical results of deterministic enzyme kinetics. 

It is clear from the previous two sections that the turnover kinetics
of single-enzyme networks is a renewal stochastic process with statistically
independent waiting times, and thus no memory. For replica networks,
there exist an initial transient regime with memory and a terminal
steady-state without it. The switch from a non-renewal to renewal
statistics in replica networks, with increasing turnover number, thus
marks a crossover from transient to steady-state regime. In the steady-state,
since waiting times are statistically independent and the governing
statistics is renewal, below we use the results of the renewal theorems
to formally link the statistical measures of counting and point processes
\cite{key-23,key-37,key-38}.

In the steady state, statistical measures at each site are related
to a pooled output, comprising of independent and identically distributed
(\emph{iid}) random variables. For counting process description, the\emph{
iid} random variables are the number of products formed at each site,
resulting in a pooled output $n=n_{1}+n_{2}+\ldots n_{\nu}$ of $n$
total number of products formed at $\nu$ sites. From this, it follows
that  $V_{ss}^{(\nu)}=\nu V_{ss}$, where $V_{ss}=\lim_{t\rightarrow\infty}d_{t}\langle n_{i}\rangle$
with $i=1,2,\cdots\nu$. For point process description, the \emph{iid}
random variables are the waiting times $\tau_{p}^{(\nu)}$ between
consecutive turnovers for $\nu$ sites, the sum of which $T_{p}^{(\nu)}=\tau_{1}^{(\nu)}+\ldots+\tau_{p}^{(\nu)}$
yields a pooled output for the $p$-th turnover time $T_{p}^{(\nu)}$.
Since $\nu$ sites are independent and identically distributed, it
follows that $\langle T_{p}^{(\nu)}\rangle=p\langle\tau^{(\nu)}\rangle$
and $\langle\tau^{(\nu)}\rangle=\frac{1}{\nu}\langle\tau\rangle$. 

Further, the renewal theorem guarantees that the single-enzyme velocity
asymptotes to the inverse mean waiting time, $\lim_{t\rightarrow\infty}V(t)=V_{ss}\equiv\langle\tau\rangle^{-1}$
\cite{key-23} . This relates the statistical measures of means for
counting and point processes, $V_{ss}^{(\nu)}\equiv\nu\langle\tau\rangle^{-1}$,
for replica networks. From this it follows that the inverse mean waiting
time $\langle\tau\rangle^{-1}$ in Eq. (\ref{eq:single-site mwt})
can be identified as the single-enzyme velocity.

In the absence of temporal correlations between turnovers, the variance
of the sum is the sum of variances, $\sigma_{T_{p}^{(\nu)}}^{2}=p\sigma_{\tau^{(\nu)}}^{2}$.
The renewal theorem, then, dictates that the squared coefficient of
variation, the randomness parameter, $r^{(\nu)}=\langle(\tau^{(\nu)}-\langle\tau^{(\nu)}\rangle)^{2}\rangle/\langle\tau^{(\nu)}\rangle^{2}$
asymptotes to the Fano factor $\rho_{ss}^{(\nu)}=\lim_{t\rightarrow\infty}\langle(n^{(\nu)}-\langle n^{(\nu)}\rangle)^{2}\rangle/\langle n^{(\nu)}\rangle$
for replica networks \cite{key-23}. A special case of this for $\nu=1$
was first introduced by Block, Schnitzer and coworkers \cite{key-24}
in the context of molecular motors, where it has widespread application
\cite{key-39,key-40}.

The above results show that for replica networks in the steady-state
the description of turnovers in terms of counts, n, and waiting times,
\textgreek{t}, are asymptotically equivalent as renewal theorems guarantee
that $V_{ss}^{(\nu)}\equiv\nu\langle\tau\rangle^{-1}$ and $\rho_{ss}^{(\nu)}\equiv r^{(\nu)}$
\cite{key-23,key-37,key-38}. However, these results are not valid
for replica networks in the transient regime where the governing statistics
is non-renewal, $\langle T_{p}^{(\nu)}\rangle\neq p\langle\tau^{(\nu)}\rangle$,
and depends on the turnover number $p$ through $T_{p}^{(\nu)}$.
Moreover, the presence of correlations between waiting times clearly
suggests that statistical measures in the transient regime should
be redefined in terms of $T_{p}^{(\nu)}$, rather than $\tau_{p}^{(\nu)}$,
as the former naturally contains correlations between waiting times.
This motivates the following definitions for the turnover number dependent
enzymatic velocity \cite{key-16},

\begin{equation}
V_{p}^{(\nu)}=\frac{p}{\langle T_{p}^{(\nu)}\rangle}\label{eq:stst-measure-point-mean}
\end{equation}
and the randomness parameter associated with $T_{p}^{(\nu)}$
\begin{align}
r_{p}^{(\nu)} & =p\frac{\langle(T_{p}^{(\nu)}-\langle T_{p}^{(\nu)}\rangle)^{2}\rangle}{\langle T_{p}^{(\nu)}\rangle^{2}}\nonumber \\
 & =p\frac{{\sum_{i}^{p}\langle(\delta\tau_{i}^{(\nu)})^{2}\rangle+\sum_{i\neq j}^{p}\langle\delta\tau_{i}^{(\nu)}\delta\tau_{j}^{(\nu)}\rangle}}{\langle\sum_{i}^{p}\tau_{i}^{(\nu)}\rangle^{2}},\label{eq:stat-measure-point-fluc}
\end{align}
where $\delta\tau_{i}=\tau_{i}-\left<\tau_{i}\right>$. 

The turnover number dependent enzymatic velocity $V_{p}^{(\nu)}$
and randomness parameter $r_{p}^{(\nu)}$ provide new statistical
measures of means and fluctuations for replica networks that can be
used both in transient and steady-state regimes, simply by increasing
$p$. In Eq. (\ref{eq:stst-measure-point-mean}), the crossover from
non-renewal $\langle T_{p}^{(\nu)}\rangle\neq p\langle\tau^{(\nu)}\rangle$
to renewal $\langle T_{p}^{(\nu)}\rangle=p\langle\tau^{(\nu)}\rangle=\frac{p}{\nu}\langle\tau\rangle$
statistics with increasing $p$, guarantees that the steady-state
enzymatic velocity is asymptotically recovered, $\lim_{p\rightarrow\infty}V_{p}^{(\nu)}=\nu\langle\tau\rangle^{-1}\equiv V_{ss}^{(\nu)}$.
In Eq. (\ref{eq:stat-measure-point-fluc}), similarly, the increase
in $p$ brings about a switch from non-renewal statistics with statistically
dependent waiting times $\sum_{i\neq j}^{p}\langle\delta\tau_{i}^{(\nu)}\delta\tau_{j}^{(\nu)}\rangle\neq0$
to renewal statistics with statistically independent waiting times,
$\sum_{i\neq j}^{p}\langle\delta\tau_{i}^{(\nu)}\delta\tau_{j}^{(\nu)}\rangle=0$.
In the asymptotic limit of large $p$, thus, $r_{p}^{(\nu)}$ reduces
to the steady-state definition, $r^{(\nu)}=\langle(\delta\tau^{(\nu)})^{2}\rangle/\langle\tau^{(\nu)}\rangle^{2}$,
which is equivalent to the steady-state Fano factor $\rho_{ss}^{(\nu)}$,
as expected from the renewal theorem \cite{key-23}. 
\begin{figure*}
\centering\includegraphics[clip,width=1\textwidth]{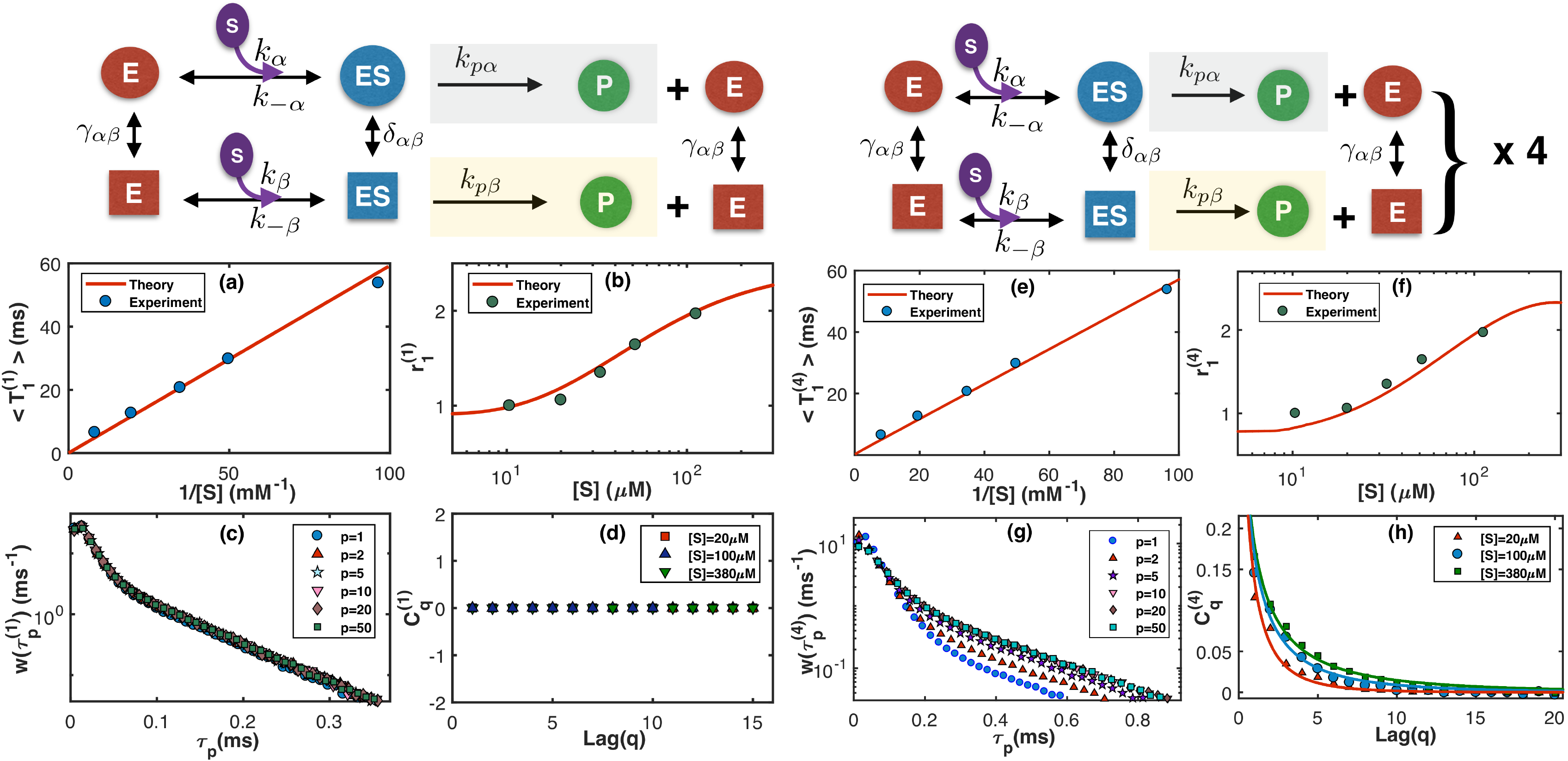}

\caption{Quantitative comparison of turnover statistics for single- and quadruple-site
ppMM models. Rate constants are estimated by \emph{simultaneously}
fitting analytical expressions for $\langle T_{1}^{(\nu)}\rangle$
and $r_{1}^{(\nu)}$ to experimental data. Exact numerical sampling
with these parameters is used to generate distributions $w(\tau_{p})$
and compute correlations $C_{q}$. The renewal character of the single-site
model and its lack of memory is confirmed in panels (c) and (d). Points
are simulation data using best-fit rate parameters and solid lines
are stretched-exponential fits $C_{q}^{(4)}=\exp[-(q/q_{0})^{\beta}]$
with $q_{0}=0.25$, $\beta=0.47,0.42,0.39$ for $[S]=20\mu\mbox{M},100\mu M,380\mu M$,
respectively. \label{fig:summary-statistics-1}}
\end{figure*}

Eqs. (\ref{eq:stst-measure-point-mean}) and (\ref{eq:stat-measure-point-fluc})
are the key results of this work as they provide the statistical measures
of point process for single-enzyme and replica networks in transient
and steady-state regimes. Their link to counting process, as shown
above, relies on the change of statistics from non-renewal at lower
$p$ to renewal at higher $p$. This naturally introduces a critical
turnover number $p^{*}$ which demarcates the transient $p\ll p^{*}$
from steady-state $p\gg p^{*}$ regime. In the steady-state, $V_{p}^{(\nu)}$asymptotes
to

\begin{equation}
V_{p\gg p^{*}}^{(\nu)}=\nu\langle\tau\rangle^{-1}\equiv V_{ss}^{(\nu)},\label{eq:c-p-mean}
\end{equation}
Similarly, $r_{p}^{(\nu)}$in the steady-state asymptotes to

\begin{equation}
r_{p\gg p^{*}}^{(\nu)}=r^{(\nu)}\equiv\rho_{ss}^{(\nu)}.\label{eq:c-p-fluc}
\end{equation}
In the transient regime, $p\ll p^{*}$, both these equivalences are
necessarily violated as the governing statistics is non-renewal. 

Eqs. (\ref{eq:c-p-mean}) and (\ref{eq:c-p-fluc}), while subsuming
the results of renewal theorems in the steady-state, provide an empirical
test of non-stationarity in experimental data and a diagnostic for
the emergence of memory in the transient regime. In the next section,
we show how these equalities can be used to determine $p^{*}$.
\begin{figure}
\centering\includegraphics[clip,scale=0.22]{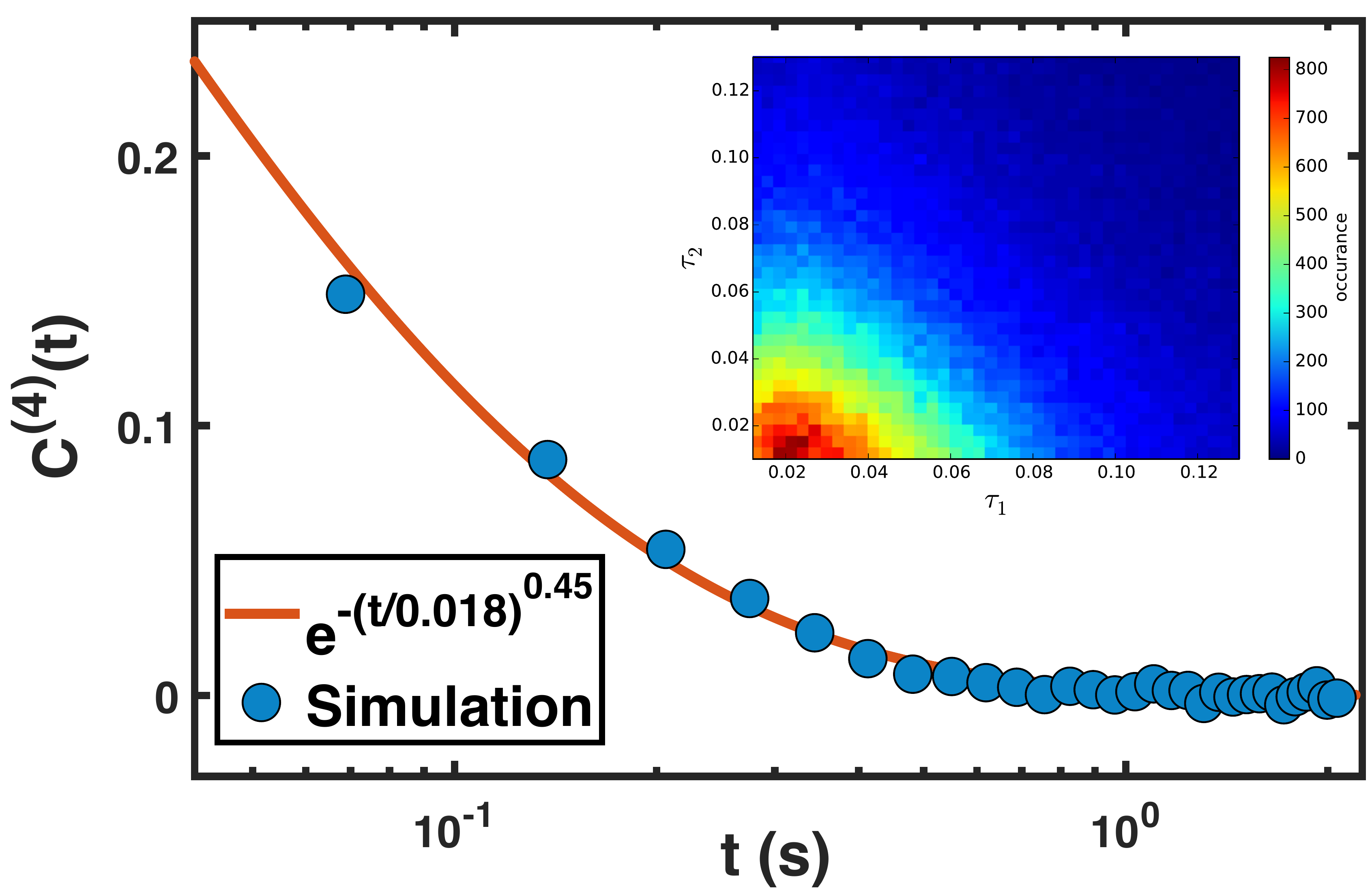}

\caption{Normalized correlation $C^{(4)}(t)$ for quadruple-site PPMM mechanism
at $[S]=100\mu M$. Points are simulation data and the solid line
is a stretched exponential $e^{-(t/t_{0})^{\beta}}$ with $t_{0}=0.018$s
and $\beta=0.45$. The inset is the joint probability $w(\tau_{1}^{(4)},\tau_{2}^{(4)})$
of consecutive waiting times, which shows that a short (long) first
waiting time is more likely to be followed by a short (long) second
waiting time, in agreement with experiment. \label{fig:Normalised-correlation-}}
\end{figure}

\section{Comparison with data\label{sec:Comparison-with-data}}

We now apply the theory developed in the preceding sections to analyze
the data from the landmark experiment in which molecular memory was
first observed \cite{key-12}. In this experiment, the catalysis of
non-fluorescent substrates to fluorescent products by the tetrameric
enzyme $\beta$-galactosidase over a range of substrate concentrations
was monitored using fluorescence spectroscopy. Waiting times were
obtained from the primary data of product turnovers as discrete fluorescence
bursts, and the distribution $w(\tau)$ and its first two moments
were computed for each substrate concentration. While the variation
of mean waiting time with $1/[S]$ was linear at low substrate concentrations,
the monotonic increase in the randomness parameter with $[S]$, bounded
below by one, was a signature of dynamic disorder. The joint distribution,
$w(\tau_{p},\tau_{p+q})$ of the waiting times, $q$ turnovers apart,
revealed that turnover events were not statistically independent but
that a short (or long) first waiting time was more likely to be followed
by another short (or long) second waiting time. This was a signature
of positive molecular memory. The correlation of waiting times, $C_{q}=\langle\delta\tau_{p}\delta\tau_{p+q}\rangle$,
remained appreciable and, when expressed in terms of a scaled time
$t=q\left<\tau\right>$, could be collapsed to a single stretched-exponential
$C(t)=\exp[-(t/t_{0})^{\beta}]$ with $\beta=0.45$ and $t_{0}=0.018$s. 

Experimental results reveal statistically dependent waiting times
and dynamic disorder in enzyme turnover kinetics of $\beta$-galactosidase.
Following the analysis of Sections \ref{sec:Renewal-processes-in}
and \ref{sec:Transients,-non-renewal-and}, this motivates us to select
the ppMM network with four replicas as the minimal model to understand
the kinetics. For comparison, we also consider a hypothetical single-enzyme
ppMM network, with identical parameters, but only a single binding
site. We use the results of Section \ref{sec:Renewal-processes-in}
and the SI to compute the marginal distribution $w(T_{1}^{(\nu)})$,
where superscript to $T_{p}^{(\nu)}$ denotes the results for a single
site $(\nu=1)$ and quadruple sites $(\nu=4).$ From $w(T_{1}^{(\nu)})$
thus obtained, we analytically compute the mean first turnover time,
$\langle T_{1}^{(\nu)}\rangle$, and the randomness parameter, $r_{1}^{(\nu)}$,
as function of the 8 rates constants. A simultaneous least-squares
fit to the experimental data provides us with the maximum-likelihood
parameters of each model. These are listed in Table \ref{tab:fitted-kinetic-parameters}
of the SI and the corresponding fits are shown in the top four panels
of Fig. (\ref{fig:summary-statistics-1}). 

The excellent agreement between model and data for both single- and
quadruple-site models leaves little to distinguish between them. Since
both models have the same number of parameters, and hence equal model
complexity, they appear to be equally plausible models for data derived
from the marginal distribution. This degeneracy in model space is
lifted by using data from the joint distribution, as we now show.
\begin{table}
\centering%
\begin{tabular}{|l|c|c|c|c|}
\hline 
Mechanism & Sites & Memory & Correlations  & $r_{1}^{(\nu)}$\emph{vs} $[S]$\tabularnewline
\hline 
\hline 
MM & 1 & Absent & $C_{q}^{(1)}=0$ & $r_{1}^{(1)}\leq1$\tabularnewline
\hline 
ppMM & 1 & Absent & $C_{q}^{(1)}=0$ & $r_{1}^{(1)}\geq1$\tabularnewline
\hline 
MM & 4 & Anti-correlated & $C_{q}^{(4)}\leq0$ & $r_{1}^{(4)}\leq1$\tabularnewline
\hline 
ppMM & 4 & Correlated & $C_{q}^{(4)}\geq0$ & $r_{1}^{(4)}\geq1$\tabularnewline
\hline 
$\beta$-galactosidase & 4 & Correlated & $C_{q}\geq0$ & $r\geq1$\tabularnewline
\hline 
\end{tabular}\caption{Qualitative comparison of turnover statistics for MM and ppMM models
for single and multiple binding sites. The ppMM model, with conformational
fluctuations and four binding sites, best agrees with the experimentally
obtained turnovers of $\beta$-galactosidase.}
\label{tab:comparison-1}
\end{table}

Since we have not found a way to obtain the joint distributions of
the Markov chain analytically, we compute them numerically from a
time series of turnovers, sampled using the Doob-Gillespie algorithm
\cite{key-41} with chain parameters set to the above least-squares
estimates. The distribution of $\tau_{p}^{(\nu)}$ and the correlation
function $C_{q}^{(\nu)}$ are shown in the bottom four panels of Fig.(\ref{fig:summary-statistics-1}). 

There is, now, a clear distinction between single-site and quadruple-site
models. The first distinction appears in the distribution of waiting
times $\tau_{p}.$ These are identically and independently distributed
for the single-site model (panels (c) and (d)) but neither identically
nor independently distributed for the quadruple-site model (panels
(g) and (h)). This confirms the results of Sections \ref{sec:Renewal-processes-in}
and \ref{sec:Transients,-non-renewal-and} that a model with fermionic
hidden states can only yield a renewal process and that multiple binding
sites are necessary for molecular memory. Focussing on panel (h),
the normalized correlation function has an excellent fit to a stretched
exponential function $C_{q}^{(4)}=\exp[-(q/q_{0})^{\beta}]$ with
parameters $q_{0}=0.25$ and $\beta=0.47,0.42,0.39$ for the substrate
concentrations $[S]=20\mu\mbox{M},100\mu M,380\mu M$ reported in
the experiment. Continuing in Fig. (\ref{fig:Normalised-correlation-})
we plot the normalized correlation function $C^{(4)}(t)$ in scaled
time $t=\langle\tau_{1}^{(4)}\rangle q$ following experiment. There
is a quantitative match between experiment and theory, with both following
a stretched exponential function $C^{(4)}(t)=\exp[-(t/t_{0})^{\beta}]$
with $t_{0}=0.018$s and $\beta=0.45$ for $[S]=100\mu$M. In addition,
the pseudocolor plot of the joint distribution of the first and second
waiting times, $w(\tau_{1}^{(4)},\tau_{2}^{(4)})$, shows that a short
(or long) first waiting time is more likely to be followed by a short
(or long) second waiting time (inset) in agreement with the molecular
memory observed in experiment. 
\begin{figure}
\centering\includegraphics[width=0.95\columnwidth]{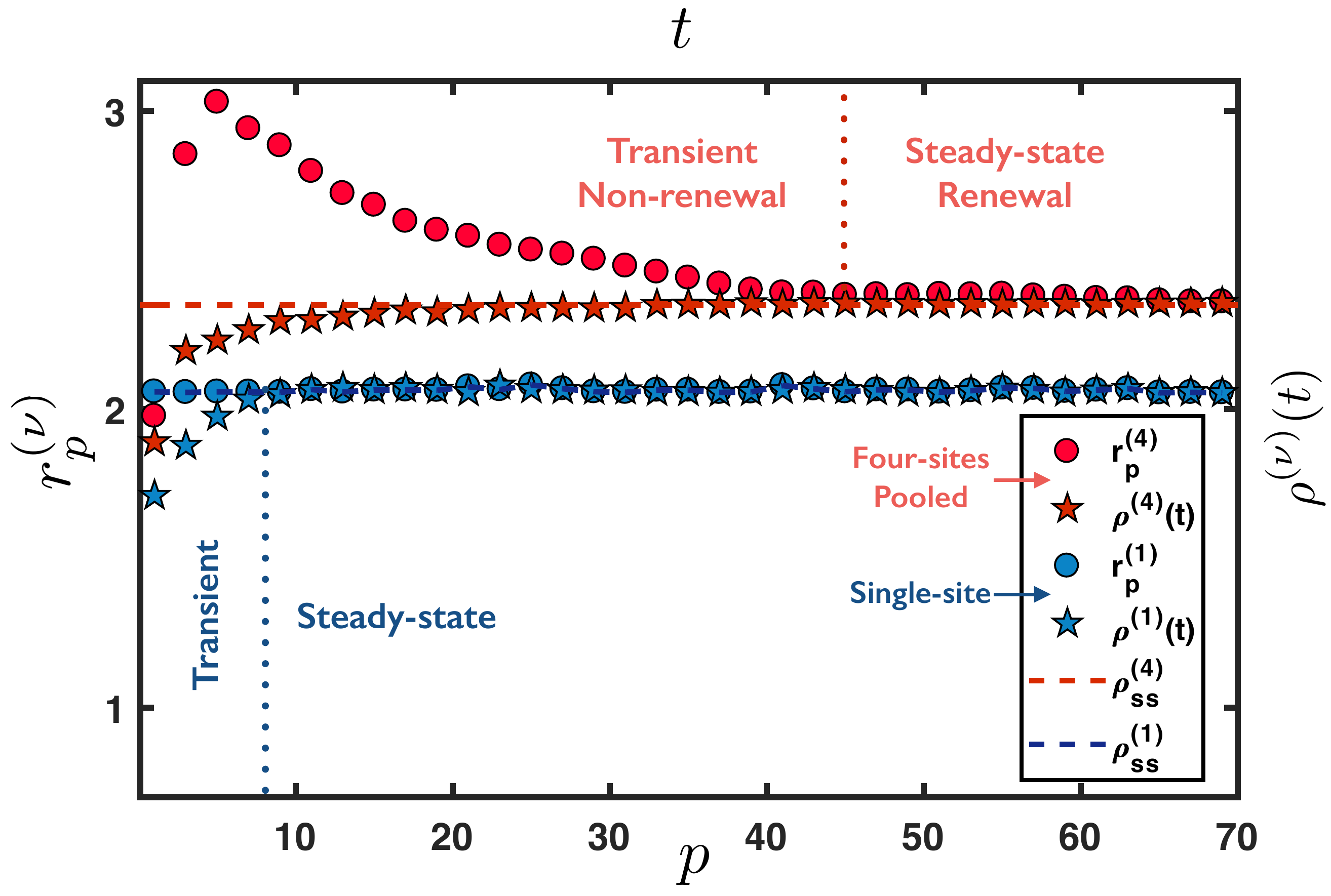}

\caption{Randomness parameters and Fano factors for the ppMM obtained from
numerically computed trajectories of the single- and quadruple-site
models. For quadruple-site model, these measures of counting and point
processes are asymptotically equivalent in the (renewal) steady state
but differ in the (non-renewal) transient state, in agreement with
Eq. (\ref{eq:c-p-fluc}). Memory persists as long as equality is violated.
For memoryless single-site model, the duration of the transient is
determined by the the counting process description, \emph{i.e.} the
time required for the Fano factor to reach its steady-state value.
Parameter values are listed in Table \ref{tab:fitted-kinetic-parameters}
of the SI\label{fig:randomness-parameter-1} }
\end{figure}

For comparison, we repeat the above calculations for the MM model
and summarize our findings in Table \ref{tab:comparison-1}. The heat
map of the joint distribution of successive waiting times, $w(\tau_{1},\tau_{2})$,
shows negative correlations for the MM model, but positive correlations
for the ppMM model. In the stationary state, the randomness parameter
is negative for the MM model, but positive for the ppMM model. Only
the ppMM model, containing both multiple binding sites and conformational
fluctuations, is in agreement with experiment, as summarized in the
last two rows of the table. While multiple sites alone yield memory,
conformational fluctuations are necessary for the correct sign of
the correlation function and the correct magnitude of the randomness
parameter. 

We now use Eqs. (\ref{eq:c-p-mean}) and (\ref{eq:c-p-fluc}) of Section
\ref{sec:Statistical-measures} to determine $p^{*}$, and thus demarcate
the transient $p\ll p^{*}$ and steady-state $p\gg p^{*}$ regimes
for the ppMM model. The variation of $V_{p}^{(\nu)}$ with substrate
concentration is shown in Fig. (\ref{fig:stat-measure-mean}) of the
SI. The enzymatic velocity in the transient regime, $V_{p\ll p^{*}}^{(\nu)}$,
deviates from the steady-state value $\nu\langle\tau\rangle^{-1}$
, where $\langle\tau\rangle^{-1}$ is the Michaelis-Menten-like (MML)
equation, Eq. (\ref{eq:ppmm-vel-ss}) in the SI. In the steady-state,
$V_{p\gg p^{*}}^{(\nu)}$ asymptotically approaches both the single-enzyme
MML equation and the classical steady-state enzymatic velocity $V_{ss}^{(\nu)}$,
in agreement with Eq. (\ref{eq:c-p-mean}). Similar analysis for the
statistical measures of fluctuations is presented in Fig. (\ref{fig:randomness-parameter-1})
, which shows the variation of $r_{p}^{(\nu)}$ with $p$ and $\rho^{(\nu)}(t)$
with $t$ for $\nu=1,4$. The comparison shows that the equivalence
between the randomness parameter and the steady-state Fano factor
is violated in the transient regime, $r_{p\ll p^{*}}^{(\nu)}\neq\rho_{ss}^{(\nu)}$,
but is asymptotically recovered in the steady-state regime $p\gg p^{*}$,
in agreement with Eq. (\ref{eq:c-p-fluc}). 

The fading of memory, the convergence of the waiting time distributions,
and the equality of the statistical measures all occur at roughly
$p^{\star}\approx50$ turnovers in this model.

\section{Conclusion and Future work\label{sec:Conclusion}}

The stochastic time-domain approach presented here, when used in combination
with the stochastic number-domain approach, namely CME, provides the
most detailed information of enzymatic reactions and can be used to
extract mechanistic information that is lost in classical deterministic
theories of enzyme kinetics.

Our work shows that the mechanism for the emergence of molecular memory
will always be operative through the transient phase in an enzyme
with multiple binding sites. It can thus be viewed as a null model
(in the sense of a null hypothesis) and should be tested against before
embarking on a search for more elaborate models of molecular memory
that may require, for instance, interactions between binding sites. 

Deviation from hyperbolicity in stochastic enzymatic networks emerges
from the concerted action of independent, and hence non-interacting,
multiple binding sites. This form of \textquotedblleft cooperativity\textquotedblright{}
is dynamic in nature \cite{key-42}, which arises from temporal correlations
between enzymatic turnovers in the transient regime, and vanishes
in the steady state regime. This contrasts the traditional description
of enzyme cooperativity in equilibrium, in which multiplicity of binding
sites and interactions between them are inextricably linked \cite{key-1}.
The replica approach presented here can be extended to include interactions
between binding sites. This can, then, pave a way to understand the
combined effect of stochasticity and interaction in generating molecular
cooperativity in the steady-state.

Our study extends beyond the simple homogeneous mechanisms studied
here, to more complex mechanisms including inhibitors \cite{key-14,key-43,key-44}
and to heterogenous catalysis of, for example, nano-particle clusters
containing numerous binding sites \cite{key-45}. In the latter case,
the method presented here can be used to estimate the catalytic rate
from turnover time data through a new kinetic measure, the heterogeneity
index, that can quantify fluctuations from non-identical binding sites
\cite{key-57,key-55,key-20}. In addition, the analysis can be extended
to unravel hidden intermediate states \cite{key-46,key-47}.

In the context of molecular motors, the renewal theorems have been
known to link the mean and variance of the physical distance moved
by the motor with the corresponding mean and variance of the number
of products formed under stationary condition \cite{key-25,key-26}.
Both are linearly proportional to time, with proportionality constants
being enzymatic velocity for the measure of mean, and diffusion coefficient
for the measure of variance. The relations between the statistical
measures of counting and point process, for the mean and variance,
presented here, and their dependence on the turnover number, can be
used to generalize the corresponding expressions for molecular motors
to non-stationary conditions, where transport coefficients are expected
to be turnover number dependent. 

Fluctuation statistics of mesoscopic quantum transport in nanoscale
devices analyzed in terms of fixed time (counting process) or fluctuating
time (point process) description using a master equation framework
\cite{key-48,key-49}. While the renewal theorems provide formal relations
between counting and point process statistics, the violation of renewal
statistics and the analysis of non-renewal fluctuation statistics,
in which temporal correlations between discrete quantum events are
accounted for, is a relatively new premise \cite{key-50,key-51}.
The generality of our results, for both renewal and non-renewal fluctuation
statistics, can provide impetus to explore a transient phase in mesoscopic
quantum transport. 

The sensitivity of the waiting time distributions to the reaction
mechanism, both in the transient and steady states, invites the application
of Bayesian probability to calculate the posterior probability $P(\mathcal{M}|\{\tau_{p}\})$,
of a model $\mathcal{M}$ given waiting time data $\{\tau_{p}\}$
and, thereon, to machine learning reaction mechanisms from turnover
data \cite{key-52,key-53,key-54}. 

To conclude, our work shows that complex forms of molecular memory
can arise from the combined action of simple memoryless steps, provides
a theoretical framework within which such action can be studied systematically,
and suggests experiments to test the validity of this generic mechanism. 

\section*{Supplementary Material}

See the supplementary information (SI) for detailed solution of the
chemical master equations for single-site and multiple-sites catalysis.

\section*{Acknowledgements}

We thank King's College for supporting A.D.'s visit to Cambridge,
where a part of this work was conceptualized. The work was first presented
as an invited lecture in the ninth conference of the Asia-Pacific
Association of Theoretical and Computational Chemistry (APATCC) at
the University of Sydney in 2019.

\section*{Data Availability}

The data that supports the findings are available within the article,
in supplemental material and in reference number {[}9{]}.

\pagebreak{}

\newpage{}

\begin{widetext}

\section*{Supplementary Information\label{sec:Supplementary-Information}}

\section*{I. Single-Site catalysis as a renewal process}

Section III in the main text has established that the turnovers for
single-site stochastic networks form renewal processes, that is, the
waiting time distributions are identically and independently distributed.
The latter has been used to show how the waiting time distribution
can be calculated from the CME describing the single-site Michaelis-Menten
(MM) network. The purpose of this section is to provide explicit results
for the waiting time distribution $w(T_{1}^{(1)})$ for the single-site
parallel pathway Michaelis-Menten (ppMM) network, from which the mean
waiting time and randomness parameter can be computed. 

\subsection{Turnover statistics of single-enzyme ppMM network}

For the parallel-pathway Michaelis-Menten model, the states are $\boldsymbol{n}=(\boldsymbol{n}^{\star},n)$,
with $\boldsymbol{n}^{\star}=(n_{E_{\alpha}},n_{ES_{\alpha}},n_{E_{\beta}},n_{ES_{\beta}})$
and the mass conservation constraint $n_{E_{\alpha}}+n_{E_{\beta}}+n_{ES_{\alpha}}+n_{ES_{\beta}}=1$.
This implies that state of the reaction at any time $t$ is of the
form $(1,0,0,0,n)$ or $(0,1,0,0,n)$ or $(0,0,1,0,n)$ or $(0,0,0,1,n)$.
Labeling these as $(E_{\alpha},n)$, $(ES_{\alpha},n)$, $(E_{\beta},n)$
and $(ES_{\beta},n)$ respectively the CME can be written as 

\begin{alignat}{1}
\partial_{t}P(E_{\alpha},n) & =k_{-\alpha}P(ES_{\alpha},n)+\gamma_{\alpha\beta}P(E_{\beta},n)-(k_{\alpha}^{\prime}+\gamma_{\alpha\beta})P(E_{\alpha},n)+k_{p\alpha}P(ES_{\alpha},n-1)\nonumber \\
\partial_{t}P(E_{\beta},n) & =k_{-\beta}P(ES_{\beta},n)+\gamma_{\alpha\beta}P(E_{\alpha},n)-(k_{\beta}^{\prime}+\gamma_{\alpha\beta})P(E_{\beta},n)+k_{p\beta}P(ES_{\beta},n-1)\nonumber \\
\partial_{t}P(ES_{\alpha},n) & =k_{\alpha}^{\prime}P(E_{\alpha},n)+\delta_{\alpha\beta}P(ES_{\beta},n)-(k_{-\alpha}+\delta_{\alpha\beta}+k_{p\alpha})P(ES_{\alpha},n)\nonumber \\
\partial_{t}P(ES_{\beta},n) & =k_{\beta}^{\prime}P(E_{\beta},n)+\delta_{\alpha\beta}P(ES_{\alpha},n)-(k_{-\beta}+\delta_{\alpha\beta}+k_{p\beta})P(ES_{\beta},n)
\end{alignat}
The analysis of Section III in the main text shows that the marginal
distribution can be written as a sum over the hidden states $\boldsymbol{n}^{\star}$.
For the ppMM network, thus, $P(n)=P(E_{\alpha},n)+P(E_{\beta},n)+P(ES_{\alpha},n)+P(ES_{\beta},n)$,
and the cumulative distribution of the turnover time is 
\begin{equation}
P(T_{p}\le t)=\sum_{n=p}^{\infty}P(E_{\alpha},n)+P(E_{\beta},n)+P(ES_{\alpha},n)+P(ES_{\beta},n)
\end{equation}
and waiting time distribution is 
\begin{equation}
w(s)=-\left[\partial_{t}P(E_{\alpha},p-1)+\partial_{t}P(E_{\beta},p-1)+\partial_{t}P(ES_{\alpha},p-1)+\partial_{t}P(ES_{\beta},p-1)\right]_{t=s}
\end{equation}
The system of differential equations must be solved with the initial
condition that, at $t=T_{p-1},$ the reaction has just entered the
state $(E_{\alpha},p-1)$ or $(E_{\beta},p-1)$. The appropriate initial
condition is $P(\boldsymbol{n})=\theta\delta_{\boldsymbol{n}\boldsymbol{m}_{1}}+(1-\theta)\delta_{\boldsymbol{n}\boldsymbol{m}_{2}}$
with $\boldsymbol{m}_{1}=(E_{\alpha},p-1)$, $\boldsymbol{m}_{2}=(E_{\beta},p-1)$
and $0\leq\theta\le1$. This gives an identical $4\times4$ system
of differential equations for the waiting times and repeating the
previous argument shows that $\tau_{p}$ are identically and independently
distributed. 

Using the notation $\boldsymbol{P}(t)=[P(E_{\alpha}),P(ES_{\alpha}),P(E_{\beta}),P(ES_{\beta})]$
and with the irrelevant index $p$ suppressed, the $4\times4$ system
of equations determining the waiting time distribution can be written
as $\partial_{t}$$\boldsymbol{P}(t)=\boldsymbol{A}\cdot\boldsymbol{P}(t)$
where
\begin{equation}
\boldsymbol{A}=\begin{bmatrix}-(k_{\alpha}^{\prime}+\gamma_{\alpha\beta}) & k_{-\alpha} & \gamma_{\alpha\beta} & 0\\
k_{\alpha}^{\prime} & -(k_{-\alpha}+\delta_{\alpha\beta}+k_{p\alpha}) & 0 & \delta_{\alpha\beta}\\
\gamma_{\alpha\beta} & 0 & -(k_{\beta}^{\prime}+\gamma_{\alpha\beta}) & k_{-\beta}\\
0 & \delta_{\alpha\beta} & k_{\beta}^{\prime} & (k_{-\beta}+\delta_{\alpha\beta}+k_{p\beta})
\end{bmatrix}
\end{equation}
The matrix exponential can be computed using the spectral representation
$\boldsymbol{A}=\text{\ensuremath{\boldsymbol{V\Lambda V}}}^{-1}$
where $\boldsymbol{V}$ is the matrix of eigenvectors of $\boldsymbol{A}$
and $\boldsymbol{\Lambda}$ is diagonal matrix of its eigenvalues.
A tedious calculation then yields

\begin{equation}
w(s)=\frac{{1}}{(A-B)(C-D)}\left[\frac{(C-D)\eta_{A}e^{-As}}{(A-C)(A-D)}-\frac{(C-D)\eta_{B}e^{-Bs}}{(B-C)(B-D)}+\frac{(A-B)\eta_{C}e^{-Cs}}{(C-A)(C-B)}-\frac{(A-B)\eta_{D}e^{-Ds}}{(D-A)(D-B)}\right]\label{eq:ppmm-wtd-ss}
\end{equation}
where where $\eta_{I}={I(k_{p\alpha}A_{1}+k_{p\beta}A_{2})-(k_{p\alpha}B_{1}+k_{p\beta}B_{2})-I^{2}k_{p\alpha}k_{\alpha}[S]}$
with $I=A,B,C,D$. In the above equation, $A$, $B$, $C$ and $D$
are the effective rate constants which are the solutions of the quartic
equation $z^{4}+\lambda_{1}z^{3}+\lambda_{2}z^{2}+\lambda_{3}z+\lambda_{4}=0$
and $A_{1}$, $A_{2}$, $B_{1}$, $B_{2}$, $\lambda_{1}$, $\lambda_{2}$,
$\lambda_{3}$ and $\lambda_{4}$, have involved dependences on the
rate constants of the ppMM model. They are given by 

\begin{eqnarray*}
{A_{1}} & = & k_{\alpha}k_{\beta}[S]^{2}+k_{\alpha}(k_{-\beta}+k_{p\beta}+{\gamma_{\alpha\beta}}+{\delta_{\alpha\beta}})[S]\\
{B_{1}} & = & k_{\alpha}k_{\beta}(k_{p\beta}+{\delta_{\alpha\beta}})[S]^{2}+(k_{\alpha}{\gamma_{\alpha\beta}}(k_{-\beta}+k_{p\beta}+{\delta_{\alpha\beta}})+{\gamma_{\alpha\beta}}{\delta_{\alpha\beta}}k_{\beta})[S]\\
{A_{2}} & = & (k_{\alpha}{\delta_{\alpha\beta}}+k_{\beta}{\gamma_{\alpha\beta}})[S]\\
{B_{2}} & = & k_{\alpha}k_{\beta}{\delta_{\alpha\beta}}[S]^{2}+(k_{\beta}{\gamma_{\alpha\beta}}(k_{-\alpha}+k_{p\alpha}+{\delta_{\alpha\beta}})+{\gamma_{\alpha\beta}}{\delta_{\alpha\beta}}k_{\alpha})[S]\\
{\lambda_{1}} & = & (k_{\alpha}+k_{\beta})[S]+k_{-\alpha}+k_{-\beta}+k_{p\alpha}+k_{p\beta}+2({\gamma_{\alpha\beta}}+{\delta_{\alpha\beta}})\\
{\lambda_{2}} & = & k_{\alpha}k_{\beta}[S]^{2}+[(k_{\alpha}k_{-\beta}+k_{\beta}k_{-\alpha})+(k_{p\alpha}+k_{p\beta}+\gamma_{\alpha\beta}+2{\delta_{\alpha\beta}})(k_{\alpha}+k_{\beta})][S]\\
 &  & +(k_{-\alpha}k_{-\beta}+k_{-\alpha}k_{p\beta}+k_{-\beta}k_{p\alpha}+k_{p\alpha}k_{p\beta})\\
 &  & +(2\gamma_{\alpha\beta}+\delta_{\alpha\beta})(k_{-\alpha}+k_{p\alpha}+k_{-\beta}+k_{p\beta})+4\gamma_{\alpha\beta}\delta_{\alpha\beta}\\
{\lambda_{3}} & = & k_{\alpha}k_{\beta}(k_{p\alpha}+k_{p\beta}+2\delta_{\alpha\beta})[S]^{2}+(k_{\alpha}[k_{p\alpha}k_{-\beta}+k_{p\alpha}k_{p\beta}+(\gamma_{\alpha\beta}+\delta_{\alpha\beta})(k_{p\alpha}\\
 &  & +k_{p\beta}+k_{-\beta})+2{\gamma_{\alpha\beta}}{\delta_{\alpha\beta}}]+k_{\beta}[k_{p\beta}k_{-\alpha}+k_{p\alpha}k_{p\beta}+(\gamma_{\alpha\beta}+\delta_{\alpha\beta})(k_{p\alpha}\\
 &  & +k_{p\beta}+k_{-\alpha})+2{\gamma_{\alpha\beta}}{\delta_{\alpha\beta}}])[S]+2\gamma_{\alpha\beta}(k_{-\alpha}k_{-\beta}+k_{-\alpha}k_{p\beta}+k_{-\beta}k_{p\alpha}+k_{p\alpha}k_{p\beta}\\
 &  & +\delta_{\alpha\beta}(k_{-\alpha}+k_{-\beta}+k_{p\alpha}+k_{p\beta}))\\
{\lambda_{4}} & = & k_{\alpha}k_{\beta}(k_{p\alpha}k_{p\beta}+\delta_{\alpha\beta}(k_{p\alpha}+k_{p\beta}))[S]^{2}+\gamma_{\alpha\beta}(k_{-\alpha}k_{\beta}k_{p\beta}+k_{-\beta}k_{\alpha}k_{p\alpha}\\
 &  & +k_{p\alpha}k_{p\beta}(k_{\alpha}+k_{\beta})+\delta_{\alpha\beta}(k_{\alpha}+k_{\beta})(k_{p\alpha}+k_{p\beta}))[S]
\end{eqnarray*}
Now $w\{s)=\mathcal{L}w(\tau)=\int_{0}^{\infty}d\tau e^{-is\tau}w(\tau)$,
where $\mathcal{L}$ is the Laplace transform operator. By corollary,

\begin{equation}
w(T_{1}^{(1)})\equiv w(\tau)=\mathcal{L}^{-1}w(s),\label{eq:wtd-ss}
\end{equation}
the explicit form for which is unwieldy and not presented here. It
can be obtained numerically from the roots of the quartic equation,
expressed above. However, the moments of $w(\tau)$ defined as $\langle\tau^{n}\rangle=\int_{0}^{\infty}d\tau\tau^{n}w(\tau)$
can be directly obtained from Eq. (\ref{eq:ppmm-wtd-ss}) using the
identity $\langle\tau^{n+1}\rangle=(-1)^{n+1}\frac{{d^{n+1}w(s)}}{ds^{n+1}}$.
The mean waiting time, $\langle\tau\rangle=-\frac{dw(s)}{ds}$, thus
simplifies to
\begin{figure}
\includegraphics[scale=0.55]{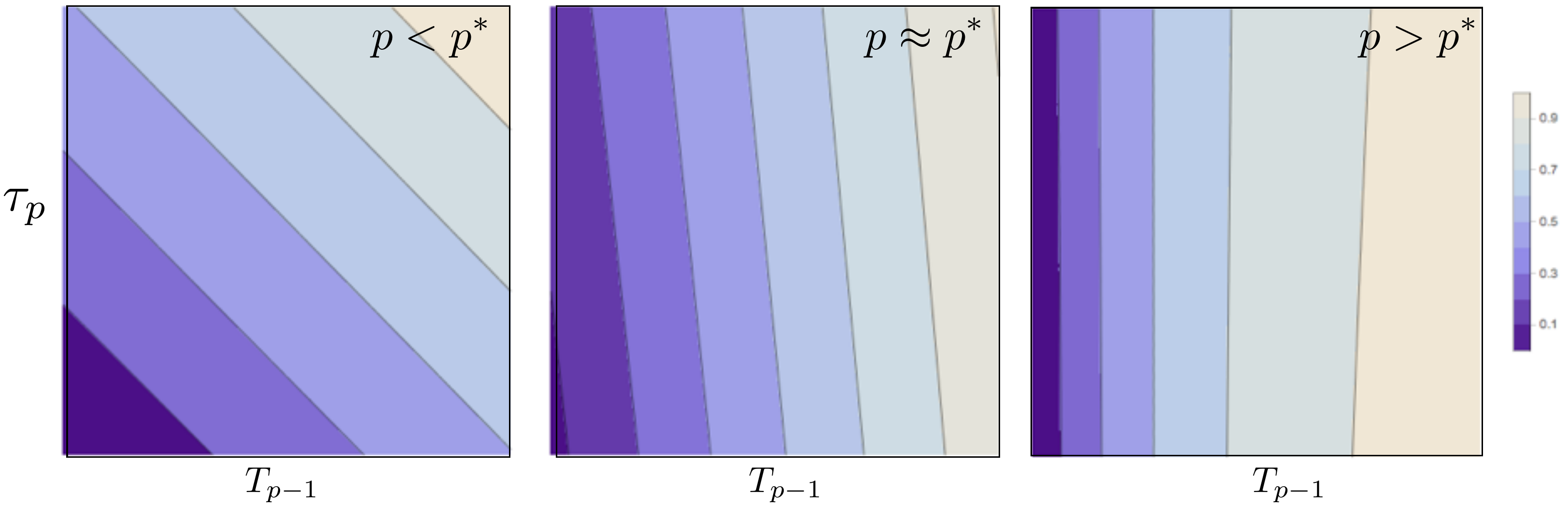}

\caption{Heat maps depicting conditional dependence of $\tau_{p}$ on $T_{p-1}$
in the transient $p<p^{*}$ crossover $p\approx p^{*}$ and steady-state
$p>p^{*}$ regimes for the MM mechanism. \label{fig:Heat-maps}}
\end{figure}

\begin{eqnarray}
\langle T_{1}^{(1)}\rangle\equiv\langle\tau\rangle & = & \frac{\lambda_{3}-k_{21}A_{1}-k_{22}A_{2}}{\lambda_{4}}\nonumber \\
 & = & \frac{E[S]^{2}+F[S]+G}{H[S]^{2}+I[S]},\label{eq:ppMM-mwt-ss}
\end{eqnarray}
the reciprocal of which yields the Michaelis-Menten-like (MML) equation
for the enzymatic velocity

\begin{equation}
v_{1}^{(1)}=\langle\tau\rangle^{-1}=\frac{H[S]^{2}+I[S]}{E[S]^{2}+F[S]+G},\label{eq:ppmm-vel-ss}
\end{equation}
where

\begin{eqnarray*}
E & = & k_{\alpha}k_{\beta}(k_{p\beta}+2\delta_{\alpha\beta})\\
F & = & k_{\beta}k_{-\alpha}k_{p\beta}+(\gamma_{\alpha\beta}+\delta_{\alpha\beta})(k_{\beta}k_{-\alpha}+k_{\alpha}k_{-\beta}+k_{\beta}k_{p\alpha}+(k_{\alpha}+k_{\beta})(k_{p\beta}+2{\gamma_{\alpha\beta}}\delta_{\alpha\beta})\\
G & = & 2{\gamma_{\alpha\beta}}(k_{-\alpha}k_{-\beta}+k_{-\alpha}k_{p\beta}+k_{-\beta}k_{p\alpha}+k_{p\alpha}k_{p\beta}+\delta_{\alpha\beta}(k_{-\alpha}+k_{-\beta}+k_{p\alpha}+k_{p\beta}))\\
H & = & k_{\alpha}k_{\beta}(k_{p\alpha}k_{p\beta}+\delta_{\alpha\beta}(k_{p\alpha}+k_{p\beta})\\
I & = & {\gamma_{\alpha\beta}}(k_{\alpha}(k_{p\alpha}k_{-\beta}+k_{p\alpha}k_{p\beta})+k_{\beta}(k_{-\alpha}k_{p\beta}+k_{p\alpha}k_{p\beta})+{\delta_{\alpha\beta}}(k_{p\alpha}+k_{p\beta})(k_{\alpha}+k_{\beta}))
\end{eqnarray*}
The randomness parameter, $r_{1}^{(1)}$, can similarly be computed
from the first and second moments of $w(T_{1}^{(1)})$.

\section*{II. Multiple-site catalysis as a non-renewal process}

The waiting time distribution for multiple-site catalysis is non-renewal
and depends on the turnover number $p$. Hence, the analysis of the
preceding section cannot be used to obtain $w(T_{p}^{(\nu)})$. The
method of Cox and Smith, based on the superposition of renewal processes,
provides a simple way to obtain $w(T_{1}^{(\nu)})$ for multiple sites
from the waiting time distribution of a single site \cite{key-36}.
In this method, the waiting time distribution for the pooled process,
\emph{i.e.} the first product turnover for catalysis at $\nu$ independent
and identical sites, can be expressed as

\begin{equation}
w(T_{1}^{(\nu)})=\nu~w(\tau)\left(\int_{\tau}^{\infty}~w(t)~dt\right)^{\nu-1}\label{eq:wtd-pp}
\end{equation}
where $w(\tau)=w(T_{1}^{(1)})$ is given by Eq. (\ref{eq:wtd-ss}).
For the ppMM network, we use Eq. (\ref{eq:ppmm-wtd-ss}) to compute
$w(T_{1}^{(\nu)})$, and thus obtain $\langle T_{1}^{(\nu)}\rangle$
and $r_{1}^{(\nu)}$ .
\begin{figure}
\includegraphics[scale=0.6]{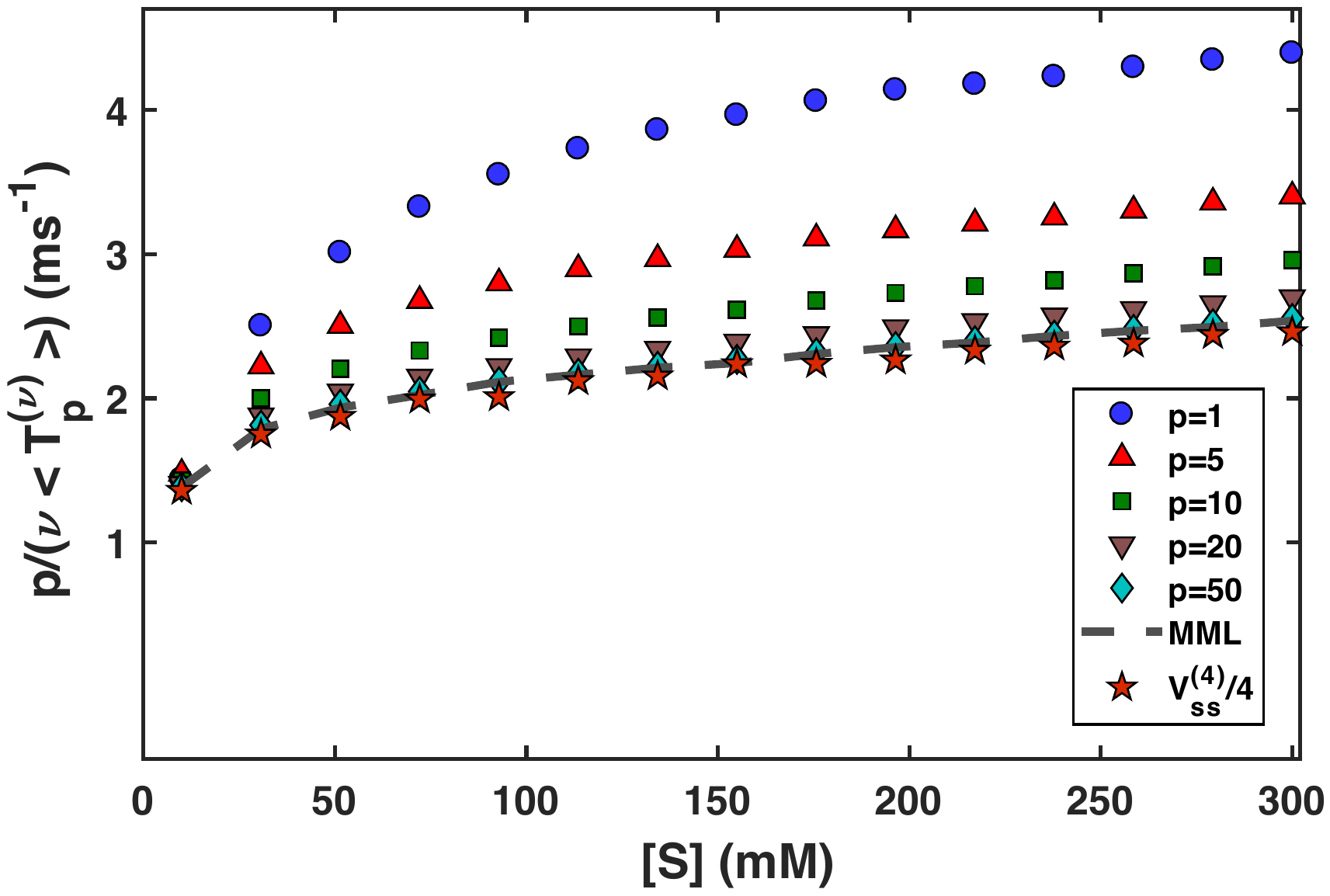}

\caption{The variation with substrate concentration, $[S]$, of the stochastically
defined enzymatic velocity, $V_{p}^{(\nu)}=p/\langle T_{p}^{(\nu)}\rangle$,
for product turnovers $p=1,5,10,20,50$, for the quadruple-site model.
The enzymatic velocity converges to its steady-state value with the
increase in $p$. This steady-state value is $\nu\langle\tau\rangle^{-1}$,
where $\langle\tau\rangle^{-1}$ is the single-enzyme velocity $v_{1}^{(1)}$,
given by the Michaelis-Menten-like (MML) equation, Eq. (\ref{eq:ppmm-vel-ss})
, shown as the dashed line. These agree with the classical definition
of the steady-state enzymatic velocity $V_{ss}^{(4)}$ plotted with
$\star$. Parameter values are listed in Table \label{fig:stat-measure-mean}.\ref{tab:fitted-kinetic-parameters}}
\end{figure}
\begin{table*}
\begin{tabular}{|c|c|c|c|c|c|c|c|c|c|}
\hline 
Model & Sites & $k_{\alpha}$ ($M^{-1}s^{-1}$) & $k_{-\alpha}$ ($s^{-1})$ & $k_{\beta}$ ($M^{-1}s^{-1}$) & $k_{-\beta}$ ($s^{-1})$ & $k_{p\alpha}$ ($s^{-1})$ & $k_{p\beta}$ ($s^{-1})$ & $\gamma_{\alpha\beta}$ ($s^{-1})$ & $\delta_{\alpha\beta}$ ($s^{-1})$\tabularnewline
\hline 
\hline 
ppMM  & 1 & $2.2\times10^{6}$ & $6.0$ & $2.2\times10^{6}$ & $6.0$ & $60.3$ & $2.9$ & $7.5$ & $16.0$\tabularnewline
\hline 
ppMM  & 4 & $5.6\times10^{5}$ & $0.4$ & $5.0\times10^{5}$ & $0.4$ & $14.6$ & $0.01$ & $1.6\times10^{2}$ & $1.05$\tabularnewline
\hline 
\end{tabular}\caption{Model parameters estimated from a simultaneous least-squares fit of
$\langle T_{1}^{(\nu)}\rangle$ and $r_{1}^{(\nu)}$ to experimental
data \cite{key-12}. }
\label{tab:fitted-kinetic-parameters}
\end{table*}

To compute the $p$-th dependent marginal and joint distributions
of turnover and waiting times, we carry out stochastic simulations.
We sample $10^{6}$ stochastic trajectories of the Markov chain for
MM and ppMM networks, using the Doob-Gillespie algorithm \cite{key-41},
and obtain a time series of turnover times $T_{p}^{(\nu)}$ and waiting
times $\tau_{p}^{(\nu)}$. We normalize histograms of these quantities
to obtain the marginal distributions $w(T_{p}^{(\nu)})$ and $w(\tau_{p}^{(\nu)})$
and their summary statistics. The trajectories are also used to compute
the number of products $n^{(\nu)}$ formed at time $t$ and their
distribution $w(n^{(\nu)},t)$. The mean and variance of the number
distribution yield the mean number of products formed $\left<n^{(\nu)}(t)\right>$
in time $t$, its rate of change $\frac{d\langle n^{(\nu)}(t)\rangle}{dt}=\lim_{\Delta t\rightarrow0}\frac{\langle n^{(\nu)}(t+\Delta t)\rangle-\langle n^{(\nu)}(t)\rangle}{\Delta t}$,
and the Fano factor $\rho^{(\nu)}(t)=\frac{\langle(n^{(\nu)}(t)-\langle n^{(\nu)}(t)\rangle)^{2}\rangle}{\langle n^{(\nu)(}(t)\rangle}$.
The joint distribution of the $p$-th and $(p+q)$-th waiting times,
$w(\tau_{p}^{(\nu)},\tau_{p+q}^{(\nu)})$ is similarly calculated
from the ensemble of trajectories. The latter is used to obtain the
normalized waiting time correlations, $C_{q}^{(\nu)}=\frac{\left<\delta\tau_{p}^{(\nu)}\delta\tau_{p+q}^{(\nu)}\right>}{\sqrt{\langle\tau_{p}^{(\nu)}\tau_{p}^{(\nu)}\rangle}\sqrt{\langle\tau_{p+q}^{(\nu)}\tau_{p+q}^{(\nu)}\rangle}}$
, where $p,q=1,2,3,\cdots$. 

\end{widetext}
\end{document}